\documentclass[acmsmall, nonacm]{acmart}
\usepackage[T1]{fontenc}
\usepackage{url}
\usepackage{amsmath}
\usepackage{amsfonts}

\usepackage{amssymb}
\usepackage{algpseudocode}
\usepackage{algorithm}
\usepackage{enumitem}
\usepackage{subcaption}
\AtBeginDocument{%
  \providecommand\BibTeX{{%
    Bib\TeX}}}

\begin{document}

\title{A Comprehensive Analysis of Accuracy and Robustness in Quantum Neural Networks}

\author{Ban Q. Tran}
\orcid{0000-0002-4100-5217}
\affiliation{%
  \institution{Department of Computer Science, Texas Tech University}
  \city{Lubbock}
  \state{Texas}
  \country{USA}}
  \email{bantran@ttu.edu}
\affiliation{%
  \institution{Department of Computing Fundamentals, FPT University}
  \city{Hanoi}
  \country{Vietnam}}
\email{bantq3@fe.edu.vn}

\author{Duong M. Chu}
\orcid{0009-0008-0806-4304}
\affiliation{%
  \institution{SAP Innovation Lab, FPT University}
  \city{hanoi}
  \country{Vietnam}}
\email{duongcmhe182219@fpt.edu.vn}

\author{Hai T.D. Pham}
\orcid{0009-0003-0976-260X}
\affiliation{%
  \institution{SAP Innovation Lab, FPT University}
  \city{hanoi}
  \country{Vietnam}}
\email{haiptd05.fpt.edu@gmail.com}

\author{Viet Q. Nguyen}
\orcid{0009-0004-1509-713X}
\affiliation{%
  \institution{SAP Innovation Lab, FPT University}
  \city{hanoi}
  \country{Vietnam}}
\email{vietnqhe190994.fpt.edu.vn@gmail.com}

\author{Quan A. Pham}
\orcid{0009-0003-3663-7572}
\affiliation{%
  \institution{SAP Innovation Lab, FPT University}
  \city{hanoi}
  \country{Vietnam}}
\email{quanpahe191162.fpt.edu@gmail.com}

\author{Susan Mengel}
\orcid{0000-0003-0722-1941}
\affiliation{%
  \institution{Department of Computer Science, Texas Tech University}
  \city{Lubbock}
  \country{USA}}
\email{susan.mengel@ttu.edu}

\renewcommand{\shortauthors}{Tran et al.}

\begin{abstract}
  Quantum Machine Learning (QML) has recently emerged as a highly promising research frontier. Within this domain, Quantum Neural Networks (QNNs),characterized by Variational Quantum Circuits (VQCs) at their core and featuring layers of quantum gates optimized by classical algorithms, have garnered significant attention. However, a rigorous and exhaustive evaluation of their practical performance remains largely incomplete. In this study, we conduct a comprehensive comparative analysis of three prominent hybrid classical-quantum architectures: Quantum Convolutional Neural Networks (QCNN), Quantum Recurrent Neural Networks (QRNN), and Quantum Vision Transformers (QViT), focusing on the critical dimensions of generalization, accuracy, and robustness. Our findings provide novel insights that address previous evaluative gaps. Notably, while these models exhibit exceptional performance on low-feature datasets such as MNIST, their learning efficacy degrades significantly when transitioned to high-feature datasets. Furthermore, convolutional-based models like QCNN appear less effective on high-dimensional data than other machine learning architectures. Additionally, while all models are susceptible to adversarial noise, traditional architectures, such as recurrent and convolutional networks, demonstrate superior resilience. Conversely, in the presence of quantum noise, the transformer-based architecture proves its strength by maintaining high robustness against measurement noise, channel noise, and finite-shot effects, whereas other architectures suffer marked performance declines. These results provide a granular perspective on the current state of the field and underscore the critical importance of tailoring model selection to the constraints of contemporary Noisy Intermediate-Scale Quantum (NISQ) environments.
\end{abstract}





\maketitle

\section{Introduction}
Quantum Machine Learning \cite{b1} has recently become quite popular in machine learning due to exhibiting superiority of quantum computing over classical computing. The core component of QML, quantum deep learning (QDL) models \cite{b4} or mainly QNNs that are constructed from parameterized quantum circuits (PQCs) \cite{b5}, have demonstrated effectiveness for downstream tasks in various domains, including computer vision, natural language processing, graphs, and more \cite{b2}. Research on QDL and QNNs has demonstrated that these models can achieve very high accuracy with a low amount of training data, significantly less than comparable classical deep learning models \cite{b15}. However, the evaluation of the generalization errors and accuracy of QNN models is not thoroughly studied. Furthermore, the resilience of quantum models against external attacks through intentional noise and quantum noise remains an issue that requires further thorough investigation. In this research, we conduct a comprehensive comparative analysis and assessment of QNN models, providing a complete perspective on the capabilities and characteristics of QNN models, including but not limited to generalization, error rate, accuracy, adversarial robustness, and average fidelity.

Previous research implementations have demonstrated the potential of QNN models to surpass classical neural networks by achieving notably impressive accuracy \cite{b10,b15,tran2025quantum}. However, the vast majority of existing studies evaluate this quantum advantage only on limited, custom-made datasets or low-feature benchmarks, such as MNIST and Fashion-MNIST. Consequently, this research provides readers with a more comprehensive perspective on the true performance of hybrid classical-quantum neural networks when applied to high-feature datasets. Furthermore, recent findings suggest that the most effective quantum deep learning models currently employ hybrid architectures \cite{bowles2024better}, so we have selected three prominent QNN frameworks, QCNN, QViT, and QRNN, for our experiments. We seek to test the accuracy and generalization of these models rigorously when scaled beyond standard input conditions, while also investigating the impact of adversarial attacks and quantum noise on robustness and in the context of real-world quantum hardware deployment. 


Our research demonstrates that QNNs, specifically Hybrid Classical-Quantum Neural Networks (HCQNNs), exhibit excellent performance on low-feature datasets like MNIST but suffer significant performance degradation when trained on high-feature datasets, such as CIFAR-10. QViT achieves high accuracy; however, this accuracy comes at the cost of a large number of trainable parameters and a significantly higher Lipschitz constant than QRNN and QCNN. For high-feature data, QRNN slightly outperforms QCNN in terms of learning outcomes. Regarding generalization error, QCNN and QRNN show a consistent trend: error scales with training set size, whereas QViT fails to exhibit this behavior due to a divergence between practical and theoretical bounds. In terms of robustness, QRNN and QCNN exhibit strong resilience against adversarial attacks, while QViT remains susceptible to such interference. Interestingly, under quantum noise conditions QViT is the least affected model, whereas QCNN is heavily impacted by Amplitude Damping, and QRNN shows diminished performance across all four noise types: Depolarizing, Bit Flip, Phase Flip, and Amplitude Damping.


The article's content is organized as follows. Section 2 presents Related Work, while Section 3 explains Theoretical Background. The parameter setup and experimental results are described in Section 4. The conclusion (5) is given in the final section.

\section{Related Works and Motivation}

A significant body of prior research has focused on evaluating QNN models, which can be categorized broadly into two types: theoretical assessments and empirical evaluations. On the theoretical front, several survey and review papers have been published to provide readers with an in-depth understanding of the fundamental nature of these models and the inherent challenges they face in practical applications. Several studies have introduced the fundamentals of QML, providing detailed explanations of its underlying mechanisms and offering theoretical justifications for why QML algorithms can achieve quantum speedup over their classical counterparts \cite{b1,b4,schuld2015introduction,b29,dunjko2018machine,ciliberto2018quantum,schuld2019quantum,huang2021power,abbas2021power}. These studies primarily consist of theoretical surveys and reviews that explore the fundamental nature and core components of current models. They provide critical discussions on the strengths and weaknesses of various architectures, as well as the inherent challenges facing QML and QNNs within the constraints of the contemporary NISQ era.

On the other hand, empirical evaluations remain relatively scarce, with only a limited number of studies providing detailed comparisons between QNNs and classical neural networks (CLNN) regarding specific properties \cite{b49,bowles2024better,ahmed2025comparative,zaman2024comparative,basilewitsch2025quantum}. Furthermore, comprehensive benchmarks that assess multiple criteria simultaneously remain scarce. During this research, we identified several studies that align with our current direction, which we use as a baseline for our evaluation. Bowles \textit{et al.} \cite{bowles2024better} conducted extensive experiments on twelve popular quantum machine learning models across six binary classification tasks. Their findings concluded that quantum models have yet to demonstrate absolute superiority over their classical counterparts across all evaluated dimensions. More detailed experiments, such as those conducted by I-Chung Chen \textit{et al.} \cite{b49}, empirically tested two quantum models—quantum self-attention and QRNN—across natural language processing (NLP) and image processing tasks. Their study included a rigorous comparison between the quantum versions and their classical counterparts, ultimately demonstrating that the quantum-enhanced models achieved superior performance. Another comparative study evaluated the performance of randomized classical networks versus QNNs, alongside a direct comparison between CNNs and QCNNs \cite{basilewitsch2025quantum}, on supervised binary image classification tasks using the MNIST dataset. Their statistical results did not indicate any clear evidence of absolute superiority for any specific model. Furthermore, the study revealed that the degree of gate entanglement and quantum correlation does not necessarily correlate in a proportional manner with the overall performance of QNN models. 

Based on the conclusions from the aforementioned studies, we recognize the critical need to \textit{conduct a rigorous performance evaluation of various QNN models to ascertain their true capabilities}. In our review, we identified two specific studies that share a common research vision and methodological alignment with our current approach \cite{ahmed2025comparative,zaman2024comparative}. These two studies, conducted by the same research group, evaluate three QNN models—Quanvolutional Neural Networks (QuanNN), QCNN, and Quantum Transfer Learning (QTL)—on only the MNIST dataset to assess their accuracy and robustness to quantum noise. However, we argue that comparing QuanNN with QCNN is conceptually inconsistent. By design, QuanNN serves merely as an image preprocessing stage through a single 'quanvolutional' layer, followed by traditional classical convolutional layers. In contrast, QCNN utilizes quantum circuits directly for its convolutional operations. Furthermore, QTL relies on transferring knowledge from a pre-trained Classical Neural Network (CLNN) to a QNN, thereby leveraging classical learning capabilities rather than inherent quantum learning. Consequently, \textit{our research fundamentally differs; we focus exclusively on evaluating the efficacy of purely quantum-circuit architectures for data learning}. While we do use classical optimizers, we made this strategic choice to mitigate the barren-plateau issue typically induced by excessive quantum circuit depth. Moreover, given that previous comparative analyses have established a thorough understanding of accuracy and quantum-noise sensitivity in QNNs, \textit{our work prioritizes characterizing adversarial perturbations in selected HCQNNs to bridge the existing knowledge gap}.

\section{Main Contributions}

The main contributions of this work are summarized across the following key areas:
\begin{itemize}
    \item \textbf{Empirical performance analysis of VQC-based hybrid classical-quantum neural networks:} We have meticulously selected three representative HCQNN architectures, QCNN, QRNN, and QViT, to conduct a comprehensive evaluation of their practical performance.
    \item \textbf{Evaluation of both accuracy and robustness:} We provide a comprehensive assessment of the most critical dimensions of QNN architectures, with a specific focus on generalization, accuracy, and robustness.To evaluate learning efficacy, we selected key metrics including accuracy and loss-based generalization error. Furthermore, to assess robustness, we utilized the Lipschitz bound and average fidelity. Collectively, these metrics provide a comprehensive representation of the overall performance and operational stability of the evaluated QNN models.
    \item \textbf{A comprehensive evaluation across both classical and quantum metrics:} Our evaluation framework incorporates both classical and quantum-specific metrics. In particular, quantum metrics, such as average fidelity, provide profound insights into the behavior of the quantum state within the Hilbert space. Regarding robustness, we conducted extensive experiments to identify and select the most impactful adversarial examples and quantum noise to measure this property rigorously.
    \item \textbf{Benchmarking the models' performance against complex, high-dimensional datasets:} To ensure a robust assessment, our experimental setup spans a diverse range of data complexities, from low-feature to high-feature environments. The resulting models were evaluated using standard benchmarks, such as MNIST and CIFAR-10, for baseline testing to analyze their efficacy in more complex, multi-channel image processing tasks.
\end{itemize}

\section{Theoretical Background}

This section presents the theoretical foundation of the QNN models used in this research. We selected three popular QNN architectures, QCNN \cite{b11}, QRNN \cite{b12}, and QViT \cite{b13}, for experimentation and comparison. We also chose several key metrics to evaluate the quality of these models, including accuracy, generalization bound, error rate, and robustness to various forms of adversarial attacks.

\begin{figure}[htbp]
\centerline{\includegraphics[scale=0.26]{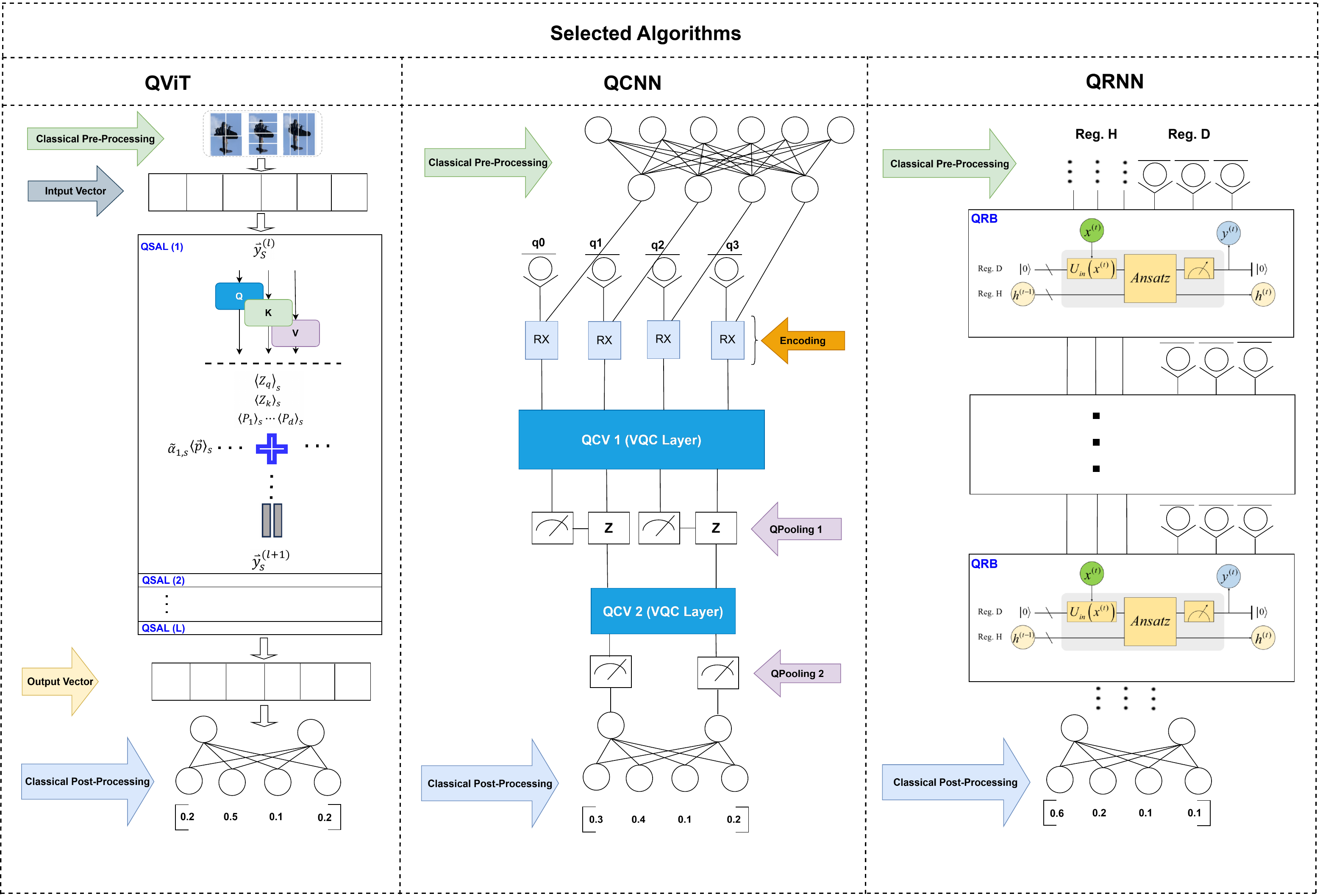}}
\Description{}
\caption{The overall architecture of the hybrid classical-quantum neural networks utilized in this study. The leftmost column presents the QViT architecture, featuring a QSAL core composed of Q, K, and V vectors, along with a Gaussian-projection self-attention mechanism. The middle column illustrates the QCNN, which leverages an inverse MERA architecture. The core of this model comprises quantum convolutional (QCV) layers and quantum pooling layers. Finally, the rightmost column depicts the QRNN model, characterized by a staggered architecture with the core QRB.}
\label{fig:QCAM}
\end{figure}

\subsection{Quantum Convolutional Neural Networks}

\begin{algorithm}[t]
\small
\caption{Training Algorithm for the Quantum Convolutional Neural Networks}
\label{alg:QCNN} 

\begin{algorithmic}[1]
\State \textbf{Input:}
\Statex \quad MNIST and Cifar-10

\State \textbf{Initialization:}
\State Import necessaries libraries and modules
\State Load and preprocess the dataset
\State QCNN architecture
\State Optimizer and learning-rate
\While{not converged}
\State \textbf{Forward evaluation:}
\ForAll{$(x, y) \in (X_{\text{train}}, y_{\text{train}})$}
    \State Encode $x$ into a quantum state
    \State Apply QCNN convolution blocks parameterized by $\theta$
    \State Apply QCNN pooling blocks parameterized by $\theta$ (measurements and downsampling)
    \State Apply final trainable unitary/classifier parameterized by $\theta$
    \State Measure output probabilities $p(y \mid x, \theta)$
\EndFor

    \State \textbf{Loss computation:} compute training loss $L(\theta)$ from predictions and labels
    \State \textbf{Gradient computation:} compute gradients $g \leftarrow \nabla_{\theta} L(\theta)$
    \State \textbf{Parameter update:} update $(\theta, S)$ using Adam optimizer with state $S$ and gradients $g$
\EndWhile

\State \textbf{Output:}
\State Store trained parameters and the model

\end{algorithmic}
\end{algorithm}

Inspired by classical CNNs, numerous studies have developed QCNN architectures based on quantum circuits. One of the popular and notable QCNN architectures is constructed using a sequence of unitary and isometry layers \cite{b11}. This QCNN architecture is based on a well-known framework in quantum information theory called the Multi-scale Entanglement Renormalization Ansatz (MERA) \cite{b18}. QCNNs have the same structure as MERA, but operate in the reverse direction. The input to the QCNN circuit is an unknown quantum state $\rho_{in}$, which is encoded from the 2D pixel array of the input image. The first quantum convolutional layer processes this quantum state. In the QCNN, the convolution layer is constructed based on a single quasi-local unitary operator $U_{i}$, functioning similarly to a $w \times w$ filter matrix in a classical CNN to generate a feature map layer, transforming the input data while preserving the quantum state with a finite depth. As this $U$ layer is constructed based on the MERA architecture, it comprises three types of isometric tensors: $t$, $w$, and $u$, wherein the tensor $t$ has two indices and is normalized to 1, and the tensor $u$ is a unitary gate (see details in Eqs. 1, 2, and 3). 
\begin{equation}
\left( t\right) \equiv \left( u\right) _{\mu \nu }^{\alpha \beta }| _{\alpha ,\beta =0}\sum _{\mu \nu }\left( t^{\ast }\right) _{\mu \nu }\left( t\right) =1
\label{eq_MERA_1}
\end{equation}
\begin{equation}
\left( w\right) _{\mu \nu }^{\alpha }\equiv \left( n\right) _{\mu \nu }^{\alpha \beta }| _ {\beta=0}\sum _{\mu \nu}\left( w^{\ast }\right) _{\mu \nu }^{\alpha }\left( w\right) _{\mu \nu }^{\alpha '}=\delta _{\alpha \alpha^{'} }
\label{eq_MERA_2}
\end{equation}
\begin{equation}
\sum _{\mu \nu }\left( u^{\ast }\right) _{\mu \nu }^{\alpha \beta }\left( u\right) _{\mu \nu }^{\alpha \beta}=\sum _{\mu \nu }\left( u^{\ast }\right) _{\alpha \beta}^{\mu \nu }\left( u\right) _{\alpha^{'} \beta^{'} }^{\mu \nu}=\delta _{\alpha \alpha^{'} }\delta _{\beta \beta^{'}}
\label{eq_MERA_3}
\end{equation}
These tensors form a lattice $\mathcal{L}$ that transforms the quantum state $\rho_{in}$ into a quantum state $|\Phi \rangle$. This quantum state $|\Phi \rangle$ represents the feature map state, analogous to the feature map in a classical CNN. In Pennylane \cite{b19}, it involves a harmonious combination of $U_3$ gates and $IsingXX$, $IsingYY$, and $IsingZZ$ gates. A subset of qubits is measured for the pooling layer, and the measurement outcomes determine which unitary rotations $V_{i}$ are applied to adjacent qubits. This transformation process reduces the output size of the layer by half, analogous to the max pooling layer in classical systems. The convolution and pooling layers are applied until the system size is sufficiently small to implement the fully connected layer. This layer consists of a unitary $F$ acting on the remaining qubits. The final qubits after this layer are measured multiple times (for example 1024 times) to determine the output. The circuit structure is fixed, while the unitaries are adjusted to enable the model to learn from the input data. This architecture yields an algorithmic complexity of only $O(log(N))$, corresponding to a doubly exponential reduction compared to other generic quantum circuit-based classifiers \cite{b55}. Fig. \ref{fig:QCAM} illustrates the theoretical layers of the CNN and a QCNN based on the MERA lattice and a practical QCNN implemented in Pennylane using $U3$, $IsingXX$, $IsingYY$, and $IsingZZ$ gates.

\subsection{Quantum Recurrent Neural Networks}


\begin{algorithm}[t]
\small
\caption{Training Algorithm for the Quantum Recurrent Neural Network (QRNN)}
\label{alg:QRNN}

\begin{algorithmic}[1]
\State \textbf{Input:}
\Statex \quad CIFAR-10 and MNIST

\State \textbf{Initialization:}
\State Import necessary libraries and modules
\State Load and preprocess the dataset
\State QRNN architecture (ancilla qubits $n_a$, encoding qubits $n_e$, sequence length $T$, depth $D$)
\State Optimizer and learning rate (Adam with cosine decay)
\State Initialize parameters $\theta \sim \mathcal{U}(-\pi, \pi)$

\While{not converged}
    \State \textbf{Forward evaluation:}
    \ForAll{$(x, y) \in (X_{\text{train}}, y_{\text{train}})$}
        \State Patchify and normalize input $x$ into a sequence $\{s_1, s_2, \dots, s_T\}$
        \For{each time step $t = 1, \dots, T$}
            \State Encode $s_t$ into encoding qubit(s) via $R_Y$ rotation
            \State Apply parameterized ansatz on all $n_a + n_e$ qubits:
            \Statex \qquad $R_X(\theta)\, R_Z(\theta)\, R_X(\theta)$ rotations, followed by
            \Statex \qquad $D$ layers of $\text{IsingZZ}(\theta)$ entangling and $R_Y(\theta)$ rotations
            \If{$t < T$}
                \State SWAP encoding qubit(s) with fresh qubit(s) for the next step
            \EndIf
        \EndFor
        \State Measure observable $O = \sum_{j} Z_j$ over ansatz qubits
        \State Compute prediction $\hat{y} = \sigma\!\bigl(\langle O \rangle\bigr)$
    \EndFor

    \State \textbf{Loss computation:} compute binary cross-entropy $\mathcal{L}(\theta)$ from predictions and labels
    \State \textbf{Gradient computation:} compute gradients $g \leftarrow \nabla_{\theta}\, \mathcal{L}(\theta)$
    \State \textbf{Parameter update:} update $(\theta, S)$ using Adam optimizer with state $S$ and gradients $g$
\EndWhile

\State \textbf{Output:}
\State Store trained parameters and the model

\end{algorithmic}
\end{algorithm}

Recurrent neural networks (RNNs) are among the earliest artificial neural networks and have numerous applications in processing sequential data \cite{b20}. Researchers have extensively explored and developed variants of RNNs in the quantum domain. Two prominent approaches stand out among the proposed solutions: purely quantum and hybrid classical-quantum. In the fully quantum architecture, a classical recurrent block is replaced by a quantum recurrent block (QRB), which is constructed based on a quantum neuron utilizing parameterized gates \cite{b21}. These QRBs perform non-linear transformations, where classical data is encoded into the quantum circuit using single rotation gates $R_{i}$ where $ i\in \left\{ X,Y,Z\right\}$. Subsequently, an ansatz with rotation gates $R_{j}$ where $ j\in \left\{ XX,YY,ZZ\right\}$, acting on two or more qubits, further processes the quantum state. The rotation angle parameters theta are updated through the quantum circuit to compute the loss function and enable the model to learn from the quantum data. Regarding the hybrid classical-quantum architecture, it also employs an ansatz based on PQCs, with the input to the recurrent block consisting of two registers: one for input/output and the other for storing historical information before being fed into the subsequent ansatz \cite{b22,b23}. The ansatz is similarly composed of layers of single-qubit and two-qubit gates. This architecture has two ways to construct a QRB. The first is the plain architecture, also known as the plain QRNN (pQRNN) \cite{b23}, and the second is the staggered architecture, or staggered QRNN (sQRNN) \cite{b12}. In pQRNN, the qubits are utilized and assigned simultaneously to the registers, whereas in sQRNN, they are assigned sequentially. Fig. \ref{fig:QCAM} illustrates the mechanism of the QRNN circuit, which utilizes a PQC to store memory information analogous to classical RNNs. Due to the sequential nature of the QRNN architecture, the data Reg. D first encodes each element of the sequential input into the quantum circuit. Then, Reg. D and Reg. H are entangled via a PQC to exploit the unique quantum mechanism. The first qubit of Reg. D is measured to obtain an intermediate prediction, after which Reg. D is reinitialized to the $|0\rangle$ state. Meanwhile, Reg. H directly feeds its quantum state into the next QRB, enabling the propagation of historical information throughout the sequence. Let’s delve a bit deeper into parameter learning. The outputs obtained from quantum circuit measurements are predictions at discrete time steps. These predictions are rescaled to align with the real values or labels. The formula for the rescaled predictions is as follows:
\begin{equation}
\overline{y}_{t}=y_{t}\cdot \left( x_{\max }-x_{\min }\right) +x_{\min } \label{eq_QRNN_1}
\end{equation}
where $x_{max}$ and $x_{min}$ are the maximum and minimum of input states.
To quantify the errors between the predicted and actual values, the QRNN model employs two loss functions: mean squared error ($L_{2}$ loss) and cross-entropy loss.
\begin{equation}L_{2}\left( \overrightarrow{\theta }\right) =\dfrac{1}{N}\sum ^{N}_{n=i}\sum ^{T}_{t=1}\left( \overline{y_{t}}\left( x,\overrightarrow{\theta }\right) -y_{true}\right) ^{2} \label{eq_QRNN_2}\end{equation}
Here, the parameter $\theta$ represents the trainable variables of the neural network, which are encoded as rotation angles in the ansatz. T denotes the number of time steps, and N refers to the data samples used in the batch training process. The learning mechanism underlying the QRNN circuit closely parallels that of the QCNN circuit, as outlined in the preceding section.

\subsubsection{Quantum Vision Transformer}


\begin{algorithm}[t]
\small
\caption{Training Algorithm for the Quantum Vision Transformer (QViT)}
\label{alg:QViT} 

\begin{algorithmic}[1]
\State \textbf{Input:}
\Statex \quad CIFAR-10 and MNIST

\State \textbf{Initialization:}
\State Import necessary libraries and modules
\State Load and preprocess the dataset
\State QViT architecture (patch size $P$, embedding dim $d$, qubits $n$, depth $D$, layers $L$)
\State Optimizer and learning-rate (Adam with cosine decay)
\State Initialize parameters $\theta$: patch embedding $W_e$, positional encoding $E_{\text{pos}}$, quantum weights $\theta_Q, \theta_K, \theta_V$ per layer, FFN weights, classifier $W_c$
\While{not converged}
\State \textbf{Forward evaluation:}
\ForAll{$(x, y) \in (X_{\text{train}}, y_{\text{train}})$}
    \State Split image $x$ into $S$ non-overlapping patches $\{p_i\}$ and project to $d$ dimensions: $z_i = p_i W_e + b_e$
    \State Add learnable positional encoding: $z \leftarrow z + E_{\text{pos}}$
    \For{each transformer layer $\ell = 1, \dots, L$}
        \State Apply Layer Normalization to $z$
        \State Encode each patch into a quantum state $|\psi_i\rangle$ (amplitude or angle encoding)
        \State Apply parameterized quantum circuits $U_Q(\theta_Q), U_K(\theta_K), U_V(\theta_V)$ to produce queries $q_i$, keys $k_i$, values $v_i$
        \State Compute attention weights $\alpha_{ij} = \exp(-(q_i - k_j)^2)$ and normalize
        \State Compute quantum self-attention output: $\text{QSAL}(z)_i = \sum_j \tilde{\alpha}_{ij} v_j$
        \State Apply residual connection and Layer Normalization
        \State Apply Feed-Forward Network (FFN) with residual connection
    \EndFor
    \State Apply final Layer Normalization, flatten, and compute logits: $f(x; \theta) = \text{flatten}(z^{(L)}) W_c + b_c$
    \State Measure output probabilities $p(y \mid x, \theta) = \text{softmax}(f(x; \theta))$
\EndFor

    \State \textbf{Loss computation:} compute training loss $L(\theta)$ from predictions and labels
    \State \textbf{Gradient computation:} compute gradients $g \leftarrow \nabla_{\theta} L(\theta)$
    \State \textbf{Parameter update:} update $(\theta, S)$ using Adam optimizer with state $S$ and gradients $g$
\EndWhile

\State \textbf{Output:}
\State Store trained parameters and the model

\end{algorithmic}
\end{algorithm}

A recent architecture that has recently become popular with researchers is the Transformer, with its core being the superior self-attention mechanism \cite{b24}. Transformers have demonstrated outstanding benchmarking performance in natural language processing, marked by the emergence of large language models. In the field of computer vision, Transformers also exhibit superior model quality \cite{b25}. Due to this superiority, studies have focused on exploring and applying self-attention to the field of quantum computing to combine the advantage of the model with the supremacy of quantum mechanics algorithms. Our investigation identifies two primary categories of quantum self-attention models explored in the literature: (1) purely quantum models and (2) hybrid classical-quantum models, which incorporate classical algorithms for post-processing. The purely quantum approach utilizes a quantum bit self-attention score matrix and quantum logic similarity to construct the quantum self-attention mechanism \cite{b26}. In contrast, the hybrid classical-quantum models integrate PQCs with a Gaussian projected self-attention mechanism while relying on classical optimization algorithms for weight updates \cite{b27}. In this study, we focus on experiments with the second model type, namely the hybrid classical-quantum approach, as fully quantum weight update mechanisms still face significant challenges due to the barren plateau problem when the quantum circuit depth increases. The QViT, also known as QSANN architecture, is composed of multiple Quantum Self-Attention Layers (QSALs), as illustrated in Fig. \ref{fig:QCAM}. Each QSAL produces an output $\overrightarrow{y}_{s}^{l}$, which is sequentially fed into the next QSAL. Each QSAL has two main components: a quantum circuit that processes the encoded input data and a classical post-processing unit.
The query vector $Q$ and key vector $K$ generate an expected value $\langle Z_{j=\{q,k\}}\rangle$ from the first qubit. 
\begin{equation}
\langle Z_{j}\rangle _{s}:=\langle \psi _{s}\left| U_{j}^{\dagger}(\theta _{j})ZU_{j}(\theta_{j})\right| \psi _{s}\rangle \label{eq_QT1}
\end{equation} 
Meanwhile, the output of the value vector $V$ consists of $d$ expectation values corresponding to $d$ Pauli string observables $P_{j}\in \left( I,X,Y,Z\right) ^{\otimes n}$, forming a $d$-dimensional vector $\overrightarrow{v_{s}}=\left[ \langle P_{1s}\rangle ,\langle P_{2s}\rangle ,\ldots ,\langle P_{ds}\rangle \right] ^{T}$ where $$\langle P_{j}\rangle =\langle \psi \left( \overrightarrow{y_{j}}^{\left( l\right) }\right) \left| U_{v}^{\dagger}\left( \theta _{v}\right) P_{j}U_{v}\left( \theta_{v}\right) \right| \psi \left( \overrightarrow{y_{j}}^{\left( l\right) }\right) \rangle$$ The classical post-processing component employs the Gaussian projected self-attention algorithm to compute their similarity based on the output of the quantum circuit. The normalized quantum self-attention coefficient between two input vectors, $s_{th}$ and the $j_{th}$, is given by: 
\begin{equation} \overline{\alpha }_{s,j}=\dfrac{\alpha _{s,j}}{\sum ^{S}_{m=1}\alpha _{s,m}} \label{eq_QT2}\end{equation}
The final output of QSAL, computed based on the parameters in Eq. \ref{eq_QT2}, is expressed as follows: 
\begin{equation}
y_{s}^{\left( l\right) }=y_{s}^{\left( l-1\right)} + \sum ^{S}_{j=1}\alpha _{s,j} . o_{j}\label{eq_QT3}
\end{equation}
This output is then passed to the subsequent QSAL layers to construct the self-attention mechanism across the entire Quantum Self-Attention Neural Networks (QSANN) architecture. Details regarding the mechanism by which the QViT circuit learns from input data through the adjustment of the parameter $\theta$ can be found in the section describing the QCNN.

\subsection{Metrics}

\subsubsection{Accuracy and Loss Functions}
Accuracy is one of the most commonly used metrics for evaluating the performance of deep learning models in classification tasks. It is widely applied in image and natural language processing tasks like sentiment classification \cite{b30}. The formula for accuracy is as follows:
\begin{equation} Accuracy=\dfrac{Y_{ccs}}{Y_{total}} \label{eq_acc}\end{equation}
where $Y_{ccs}$ is the correclty classified samples and $Y_{total}$ is the total number of samples. To explain more simply, using the confusion matrix for a binary classification problem, with the components true positive (TP), false positive (FP), true negative (TN), and false negative (FN), the formula for accuracy can be expressed as:
\begin{equation} Accuracy=\dfrac{TP + TN}{TP + TN + FP + FN} \label{eq_accbin}\end{equation}
There are several variations of loss function formulations used for classification problems \cite{b31,b32}. This study focuses on two primary loss functions: Binary Cross-Entropy (BCE), also known as log loss, and Categorical Cross-Entropy (CCE), known as multi-class log loss. For binary classification tasks, BCE is a standard loss function commonly used for evaluation. Suppose each label $y_i\in \left\{ 0,1\right\}$ is an element in the dataset $\left\{ \left( x_{i},y_{i}\right) \right\} _{i=1}^{n}$. We define $\widehat{p}_{i}=f_{\theta }\left( x_{i}\right)$ as the probability that the model predicts $y_{i}=1$. The BCE loss value for a single instance is computed using Eq. \ref{eq_bcesin} below.  
\begin{equation}L\left( y_{i},\widehat{p}_{i}\right) =-\left[ y\log \left( \widehat{p_{i}}\right) +\left( 1-y_{i}\right) \log \left( 1-\widehat{p}_{i}\right) \right]\label{eq_bcesin}\end{equation}
In practice, the main objective is to minimize the average loss over the entire dataset as the Eq. \ref{eq_bceover}.
\begin{equation}\mathcal{L}_{BCE}\left( \theta \right) =\dfrac{1}{n}\sum ^{n}_{i=1}L\left( y_{i},\widehat{p}_{i}\right)\label{eq_bceover}\end{equation}
CCE is an extension of BCE for multi-class classification problems. Suppose we have $C$ classes and a dataset of labels $y_{i}\in \left\{ 0,1\right\} ^{C}$. Let $\widehat{p}_{i}=\left[ \widehat{p}_{i,1},\ldots ,\widehat{p}_{i,C}\right]$ denote the predicted probability distribution over the classes for sample $i$. The loss for an individual sample is then computed as follows:
\begin{equation}L\left( y_{i},\widehat{p}_{i}\right) =-\sum ^{C}_{j=1}y_{i,j}\log \left( \widehat{p}_{i,j}\right)\label{eq_ccesin}\end{equation}
where $y_{i,j} \in \left\{ 0,1\right\}$ and $\sum _{j}y_{i,j} = 1$
The average loss over $n$ samples is obtained at the dataset level:
\begin{equation}\mathcal{L}_{CCE}\left( \theta \right) =-\dfrac{1}{n}\sum ^{n}_{i=1}\sum ^{C}_{j=1}y_{i,j}\log \left( \widehat{p}_{i,j}\right) = \dfrac{1}{n}\sum ^{n}_{i=1}L\left( y_{i},\widehat{p}_{i}\right)\label{eq_cceover}\end{equation}

\subsubsection{Generalization Bound}

In classical and quantum deep learning models, avoiding overfitting is a key objective during training. To assess whether a model can generalize well beyond the training data, a commonly used metric is the generalization bound \cite{b28,b29}. To estimate the generalization bound, one must first evaluate the generalization error, which quantifies the discrepancy between the model's empirical performance on the training dataset and its expected performance over the actual data distribution. This metric indicates the model's ability to generalize to previously unseen data and is critical for assessing robustness in classical and quantum deep learning frameworks. We now discuss the generalization property of a quantum deep learning model. Classical input data $x$ is encoded into a quantum state $\rho(x)$ via a quantum channel $\varepsilon_{\alpha}(\rho(x))$, which is parameterized by trainable parameters $\alpha$. The expected loss function, denoted as $\mathcal{R}$, is defined as follows:
\begin{equation} \mathcal{R}\left( \alpha \right) =E_{\left( x,y\right) \sim \mathcal{P}}\left[ \mathcal{L}\left( \alpha ;x,y\right) \right] \label{eq_GB1}\end{equation}
where x are features and y are the labels, and $\mathcal{P}$ is the joint distribution.
Assuming we have a dataset $S=\left\{ \left( x_{i},y_{i}\right) \right\} _{i=1}^{N}$ consisting of $N$ samples used to train a QML model, the model's performance is evaluated based on the average loss over the dataset, commonly referred to as the training loss.
\begin{equation} \widehat{\mathcal{R}}\left( \alpha \right) =\dfrac{1}{N}\sum ^{N}_{i=1} \mathcal{L}\left( \alpha ;x_{i},y_{i}\right) \label{eq_GB2}\end{equation}
The difference between the expected loss and the training loss corresponds to the value of the generalization error.
\begin{equation} GR\left( \alpha \right) =\mathcal{R}\left( \alpha \right) -\widehat{\mathcal{R}}_{s}\left( \alpha \right) \label{eq_GB3}\end{equation}
The generalization error is expected to depend on the model's richness and the amount of training data. A recent in-depth study \cite{b15} has shown that, for a QML model with $T$ parameterized local quantum channels and $N$ training samples, the generalization error exhibits the following algorithmic complexity.
\begin{equation}  GR\left( \alpha \right) \sim \mathcal{O}\left( \sqrt{\dfrac{T\log T}{N}}\right) \label{eq_GB4}\end{equation}


\subsubsection{Lipschitz Bound}
The human eye can easily distinguish a noise-injected image from a clean legitimate image. However, this ability to distinguish is not as easy for a deep learning model. Studies have shown that quantum deep learning models also struggle with adversarial examples \cite{b16}. To measure the robustness of deep learning models in this study, we use the Lipschitz bound \cite{b34}, a metric in which the model is treated as a function $f$, and the Lipschitz bound of $f$ is defined as the smallest constant $L$ in the expression:
\begin{equation}|| f\left( x\right) -f\left( y\right) \left\| \leq L\right\| x-y\|\label{eq_lip1}\end{equation}
When the Lipschitz bound is high, it indicates that a small change in the input can result in a significant change in the output, which means the model is more sensitive to noise and more susceptible to attacks as compared to models with a low Lipschitz bound value. From Eq. \ref{eq_lip1}, the Lipschitz bound value of a QNN is computed using Eq. \ref{eq_lip2} below:
\begin{equation}L_{\theta }=2\left\| M\right\| \sum ^{N}_{j=1}\left\| \omega_{j}\right\| \left\| H_{j}\right\|\label{eq_lip2}\end{equation}
where and $\omega_{j}$ is fixed and $\theta_{j}$ is a trainable parameter.

\subsubsection{Average Fidelity}
Classical input data, such as images, are encoded into quantum states before being fed into the quantum deep learning model for training. The quantum states of clean legitimate and adversarial examples after encoding may vary. To assess the model’s robustness against adversarial examples, the average fidelity metric is used to measure the difference between the quantum states before and after adding noise \cite{b35}. Eq. \ref{eq_fide1} illustrates how this metric is computed. In this expression, a trace-preserving quantum operation is denoted by $\varepsilon$, $U$ represents a unitary transformation, and $| \psi \rangle$ signifies a quantum state in a $d$-dimensional Hilbert space. The average is performed over the complete set of quantum states, which are randomly drawn uniformly following the Haar measure, where the normalization condition $\int d\psi = 1$ holds.
\begin{equation} \overline{F}\left( \varepsilon,U \right) \equiv \int d\psi \langle \psi \left| U^{\dagger} \varepsilon \left( \psi \right) U \right| \psi \rangle \label{eq_fide1}\end{equation}
Eq. \ref{eq_fide1} can be rewritten more concisely by computing the inner product and taking the norm of the encoded quantum states corresponding to the clean legitimate and adversarial example images \cite{b16}. 
\begin{equation}F= \left| \langle \psi ^{adv}\right| \psi ^{leg}\rangle | ^{2}\label{eq_fide2}\end{equation}
where $| \psi ^{adv}\rangle$ and $| \psi ^{leg}\rangle$ are the quantum states after encoding the adversarial example and clean legitimate images.

\subsubsection{Quantum Noise}

Quantum noise represents the undesirable environmental interactions that lead to decoherence, a process that degrades fragile quantum properties, such as superposition and entanglement. These external perturbations induce unintended transitions in qubit states, ultimately compromising the fidelity and accuracy of computational results. In this study, we evaluate the performance of three HCQNN models under quantum-native perturbations, including measurement noise, channel noise, and finite-shot effects. Among these various forms of quantum noise, we give particular consideration to quantum channel noise, specifically focusing on Bit-flip, Phase-flip, Phase-damping, Amplitude-damping, and Depolarization channels.


\section{Experiments Setup and Results}

In this experimental section, we aim to evaluate the capabilities of QNN models, QCNN, QRNN, and QViT, in image classification tasks, including binary and multi-class classification. Our target is to assess the ability of these models to learn from data. Specifically, we intend to determine whether they can generalize effectively or if they tend to suffer from overfitting. Additionally, we want to evaluate the practical performance of the models and provide deeper insights into their effectiveness in multi-class classification tasks.

\subsection{Dataset and Preprocessing}

Several popular and standard datasets are used for image classification tasks in deep learning, such as MNIST, Fashion-MNIST, CIFAR-10, CIFAR-100, and ImageNet. In this study, we selected MNIST \cite{b42} and CIFAR-10 \cite{b43} because these datasets have a moderate number of images, neither too large nor too small, making them suitable for our experimental configuration. These two datasets are the most widely used in previous QNN research. MNIST (Modified NIST) is a dataset with 70,000 grayscale images sized 28x28 pixels, containing handwritten digits from 0 to 9. The dataset is divided into 60,000 training images and 10,000 test images. CIFAR-10 is a dataset with 60,000 color images of 32x32 pixels, evenly distributed across 10 classes (such as planes, automobiles, birds, cats, and dogs). The dataset includes 50,000 training images and 10,000 test images. Both datasets are standard benchmarks in multi-class image classification tasks and for evaluating the generalization capability of deep learning models. MNIST, with its small size, simplicity, and low noise, is suitable for implementing quantum models with limited qubit counts. CIFAR-10, with its color characteristics and more complex features, allows us to verify the learning capability and stability under various levels of adversarial attacks.

\begin{figure}[htbp]
\centerline{\includegraphics[scale=0.5]{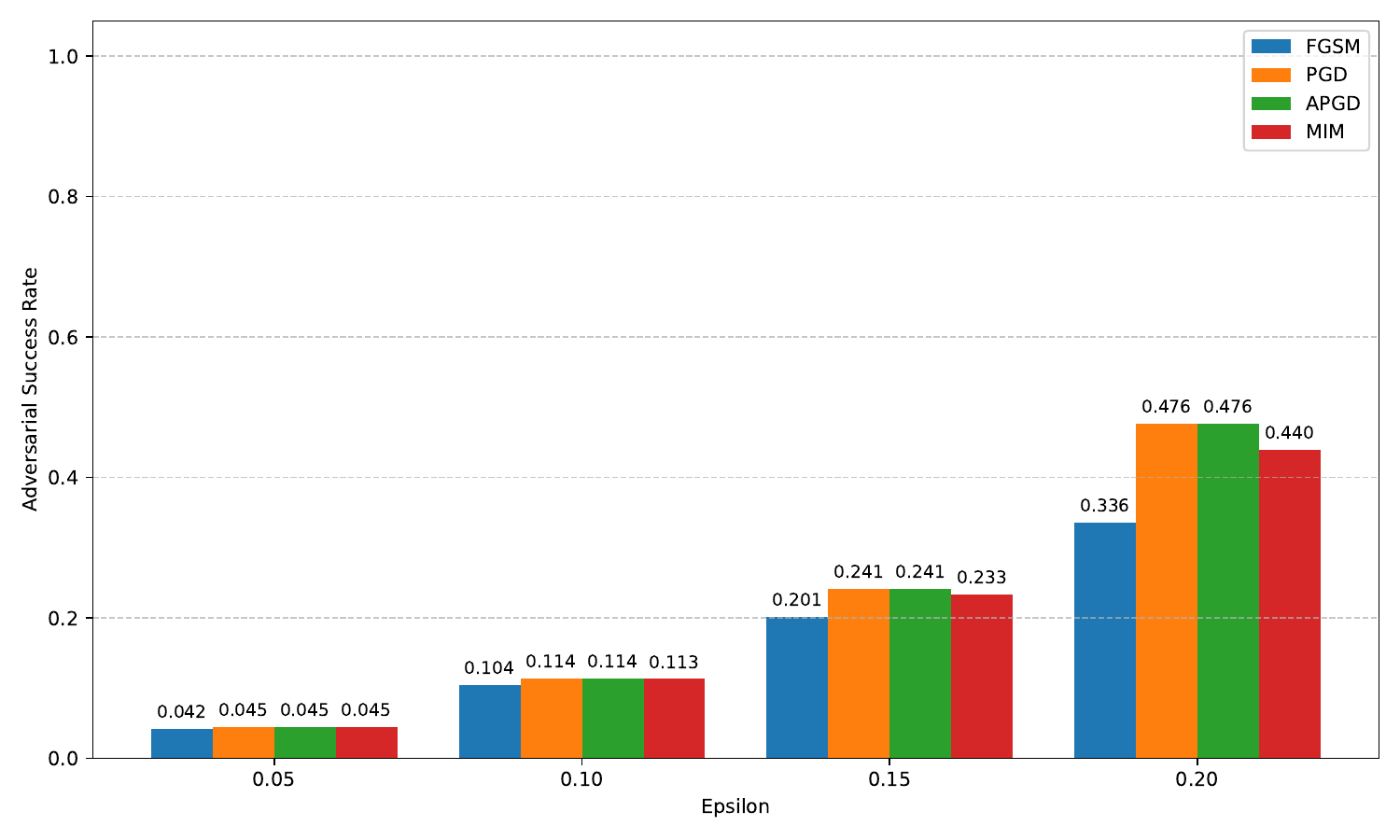}}
\Description{}
\caption{Compare the ASR of QCNN with the four main adversarial attack methods: FGSM, PGD, APGD, and MIM. PGD and APGD yielded the highest successful attack rates among the four methods.}
\label{fig:Adv}
\end{figure}


In QNN models, input data, after preprocessing, needs to be mapped into a quantum state to be compatible with the model's structure. Classical data is mapped through unitary transformations. The two encoding methods used in this experiment are amplitude encoding and angle encoding. During the experiment, three distinct data preprocessing strategies are developed to suit the specific architectures of each quantum model, including QRNN, QCNN, and QViT. For the QRNN model, images are divided into small, uniform regions, and the average value of each area is calculated to form a representative feature sequence. This sequence is then normalized and mapped to a quantum state using angle encoding, clearly reflecting the sequential nature of the quantum recurrent network. For the QCNN model, images are normalized and resized to 8x8, then flattened into a one-dimensional vector before being mapped using amplitude encoding. For the QViT model, the preprocessing involved normalizing the data, dividing the image into small patches of uniform size, flattening each patch, and then applying amplitude encoding to transform the input data into a suitable quantum state.


To test the resilience against adversarial attacks, we applied four methods for generating adversarial examples: Fast Gradient Sign method (FGSM) \cite{b44}, Projected Gradient Descent (PGD) \cite{b45}, Auto PGD (APGD) \cite{b46}, and Momentum Iterative Method (MIM) \cite{b47}. FGSM is a single-step attack method that minimizes the negative loss function to create perturbations in the input. PGD is an extension of FGSM, applying multiple steps (multi-step variant) and projecting the result back to the nearest point within a perturbation budget if the point lies outside. APGD improves PGD by adapting the step size and addressing its weaknesses such as overshooting, local minima traping, coordinate-wise variance, hyperparameter tuning dificulty, ones-ize-fits-all and so on. MIM enhances FGSM by incorporating a momentum mechanism to maximize the loss function and overcome local minima, thereby increasing attack effectiveness. All four methods are used to evaluate the robustness of the quantum models, and the testing process showed that APGD had the most significant impact on the performance of deep learning models. (See detailed results in Fig. \ref{fig:Adv}). Therefore, we selected APGD as the primary attack method to assess the resilience of quantum models under the most challenging attack conditions. In this experimental process, we used the APGD attack configuration with the following parameters: Steps/N = 100; first step size $\eta^{(0)=\epsilon/N}$; momentum $\alpha = 0.75$; and condition ratio $\rho = 0.75$; max norm ball with  radius $\epsilon \in [0.1,0.2,0.3,0.5]$. We varied the perturbation values $\epsilon$ to test the resilience of the quantum models.

\subsection{Model Setting}

In this section, we present the experimental configuration of the three QNN models along with their parameters. Regarding QViT, we use amplitude encoding because this method reduces computation time more than angle encoding. Additionally, incorporating specific components of the classical Transformer results in the QViT model having significantly more parameters than the other models. During the experiments, we observed that the model requires a certain level of complexity to perform effectively on the CIFAR-10 dataset. In contrast, the MNIST dataset does not require as many parameters to achieve good results. Therefore, compared to MNIST, we used a more complex configuration for CIFAR-10. Compared to the model proposed by Li et al. \cite{b27} and in addition to the quantum self-attention block, this model incorporates Positional Encoding, Layer Normalization, and a Position-wise Feedforward block, inherited from the classical vision transformer model \cite{b22}. This inheritance enables the model to achieve good results in image classification tasks. For the MNIST dataset, we used d\_model = 32 and iterated the encoder layer once. The more complex CIFAR dataset led us to increase d\_model to 96 and iterate the encoder layer twice. Additionally, we utilized five qubits and a circuit depth of 8 for both datasets.

For QRNN, the data is mapped using angle encoding, where each element in the sequence is directly encoded into the corresponding rotation angles. This architecture mimics the capability of classical RNNs, extended by the staggered and entanglement phenomena. Each sequence element is encoded through a $R_y$ rotation gate, followed by a sequence of single gates $R_x$, $R_y$, $R_z$, and the multi-qubit $IsingZZ$ gate, repeated twice. The number of qubits used for MNIST is eight. For CIFAR-10, the number of qubits used is 12. Similarly, QViT requires a larger number of qubits than QRNN because its attention mechanism must encode multiple image patches along with positional information in superposition. With only 15 qubits available in our setup, the model cannot represent enough patches to capture meaningful spatial dependencies even for MNIST. This limitation becomes more pronounced for CIFAR-10, where the richer image features demand both a higher qubit count and greater circuit depth to achieve competitive performance.


The QCNN model is implemented with two quantum convolutional layers, two quantum pooling layers, and a final quantum fully connected layer. After the input quantum state is prepared, each qubit is first transformed by a $U3$ gate to bring it into an appropriate basis. Subsequently, pairs of adjacent qubits undergo the sequence of $Ising$ gates, generating local entanglement and enabling information mixing between neighboring qubits. The quantum pooling layers are implemented by measuring a subset of qubits on a computational basis. The measurement outcomes control $U3$ gates acting on neighboring unmeasured qubits. This process introduces nonlinearity into the circuit while eliminating the measured qubits, thereby halving the number of remaining qubits in the system, similar to dimensionality reduction in classical pooling. After passing through two consecutive convolution-pooling blocks, the remaining qubits are fed into the quantum fully connected layer, implemented by an ArbitraryUnitary gate acting simultaneously on all final qubits. This layer aggregates all features extracted and reduced from the previous layers, performing a role analogous to the fully connected layer in classical CNNs. This model utilizes six qubits with a circuit depth of 12.  Studies have indicated that QML does not require as much input data as classical models to achieve promising generalization ability \cite{b15}. Furthermore, training QML models, when simulated on classical computers, demands significant computational resources. Therefore, to optimize resources in this research, we performed experiments only on the two datasets most frequently utilized in prior QML studies, as indicated by our statistics. To evaluate whether the encoding method affects model performance, we compared the accuracy of three models, QCNN, QRNN, and QViT, using the same amplitude encoding method on the CIFAR-10 dataset. 


\subsection{Experimental Scenarios}

\begin{figure}[htbp]
\centerline{\includegraphics[scale=0.55]{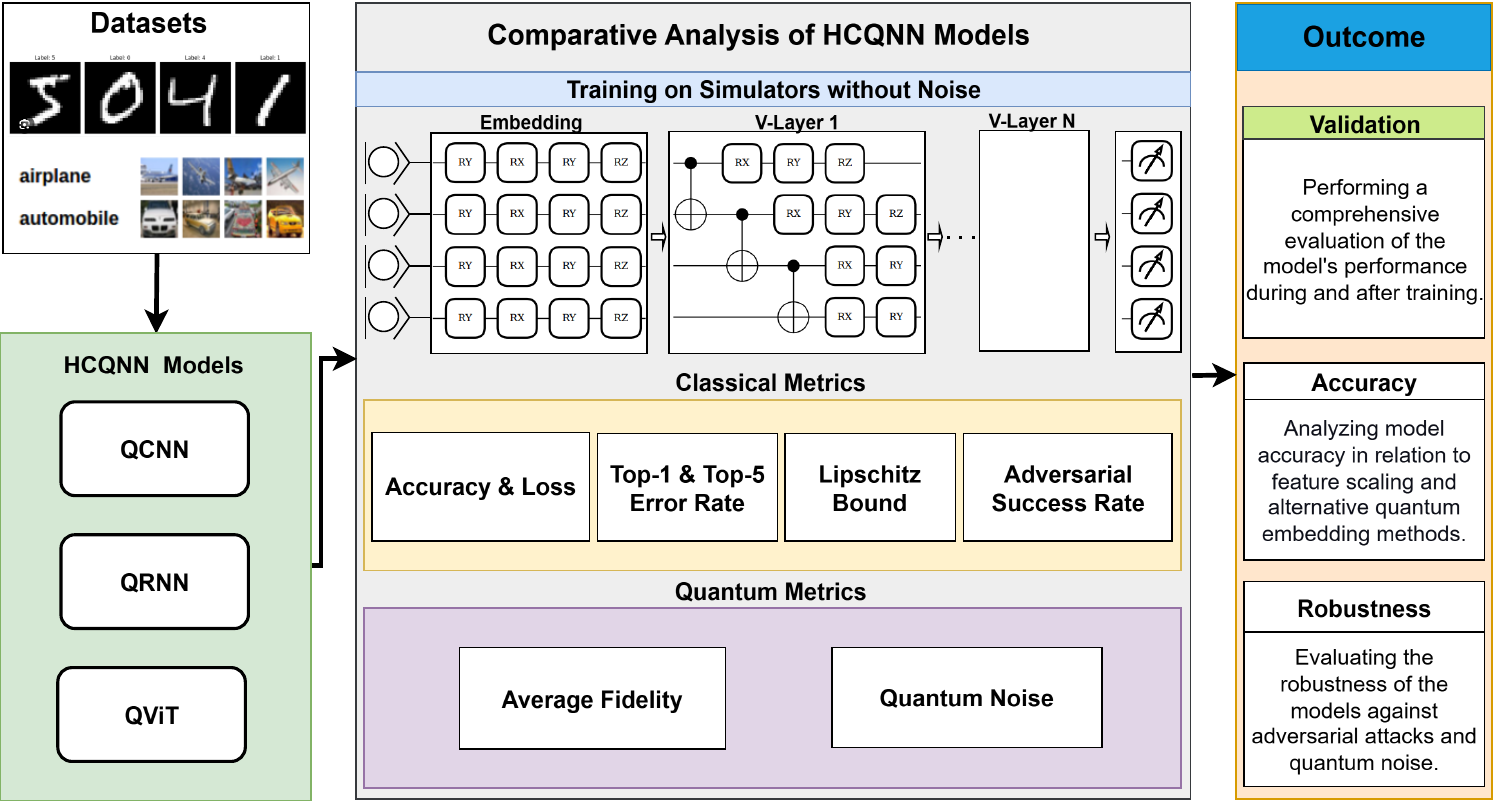}}
\Description{}
\caption{An overview of the analytical methodology and evaluation framework employed in this research.}
\label{fig:QCAM-method}
\end{figure}

Previous studies primarily focused on evaluating the generalization or robustness of individual QNN models and on comparing them with classical deep learning models. In this study, we focus on both aspects and conduct an in-depth evaluation of QNN models exclusively. We employed two experimental scenarios to conduct the research. In the first measurement scenario, we focused on evaluating the baseline performance of three models in image classification tasks without any noise or adversarial attacks. The primary objective was to assess how much these models can learn and generalize from clean data. We utilized standard image datasets for both binary and multi-class classification tasks. The experimental procedure comprised (i) data preparation, (ii) training, utilizing cross-entropy loss and monitoring metrics on the validation set, and (iii) evaluating metrics on the final test set. In conducting the above scenario, we performed experiments with different train/test sizes to evaluate the performance of various models under varying data volume conditions. Specifically, for CIFAR-10 multi-class, the training data set sizes used were (1000, 2000, 5000, 10000, 20000, 50000), and for MNIST multi-class, the train sizes were (1000, 2000, 5000, 10000, 20000, 60000). The binary CIFAR-10 pairs used were (airplane vs. automobile, automobile vs. truck, cat vs. dog), and the binary MNIST pairs used were ([0, 1], [0, 8], [1, 7], [6, 9]). All train sizes were paired with a test size, maintaining a test/train ratio o 1/5. The second scenario was designed to test the robustness of the models when confronted with adversarial attacks and quantum noise. This scenario is a critical step in evaluating the safety and resilience of QNN architectures in practical settings, where input data may be subtly altered. We applied the most disruptive attack type, APGD, through computation and comparison. The adversarial attack experimental procedure includes: (i) preparation: using fully trained models, (ii) generating adversarial examples: employing only APGD with varying parameters $\epsilon \in (0.1, 0.2, 0.3, 0.5)$, and (iii) evaluation: calculating metrics on the adversarial examples and comparing performance before and after the attack. To evaluate the performance of our models during and after training, we selected several fundamental types of quantum noise, including measurement noise, channel noise, and finite-shot effects. Among these, our primary focus is on channel noise, specifically encompassing Bit-flip, Phase-flip, Phase-damping, Amplitude-damping, and Depolarizing channels. Through these two scenarios, we aim to provide an in-depth analysis of both the learning capability and the robustness of QNN models, thereby shedding light on their role and potential.

\begin{figure}[htbp]
\centerline{\includegraphics[scale=0.4]{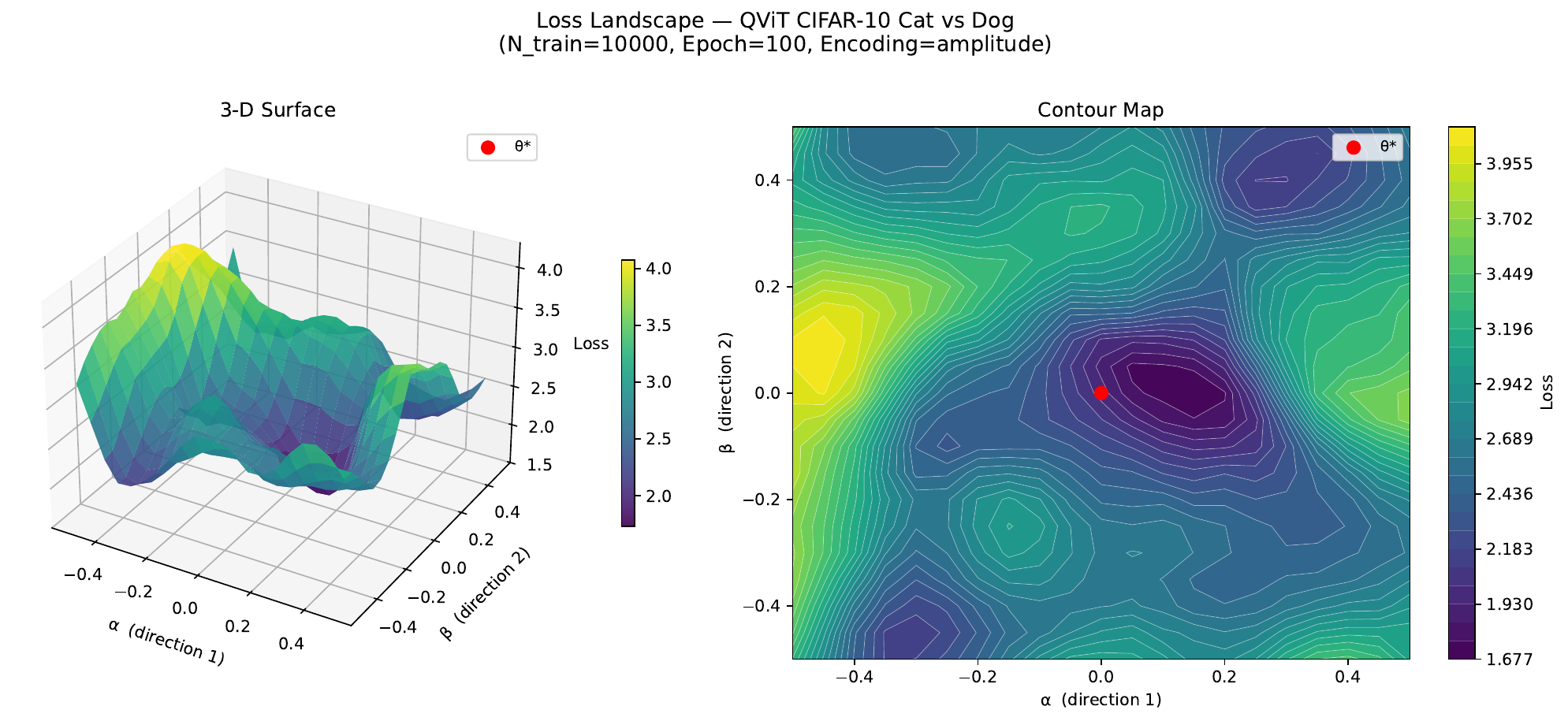}}
\Description{}
\caption{2D and 3D loss landscapes of the QViT optimizer for Cat and Dog classification with 10,000 samples, 100 epochs, and the amplitude encoding method.}
\label{fig:landscape}
\end{figure}

\subsection{Results and Analysis}

\begin{table*}[htbp]
\centering
\caption{Evaluation of the prediction accuracy of the QNN models on the CIFAR-10 (Cat vs. Dog) and the MNIST (1 vs. 7) clean dataset}
\setlength{\tabcolsep}{3pt}
\begin{tabular}{p{80pt}p{60pt}p{30pt}p{90pt}}
    \toprule
    \textbf{Model} & \textbf{Accuracy(\%)} & \textbf{Loss} & \textbf{Generalization Bound} \\
    \midrule
 QCNN (CIFAR-10) & 55.5 &  0.48  & 0.003 \\
 
 QCNN (MNIST) & 97.3 &  0.23  & 0.002 \\
 \hline
 QRNN (CIFAR-10) & 57.1 &  0.67  & 0.013 \\
 
 QRNN (MNIST) & 96.7 &  0.23 & 0.003 \\
 \hline
 QViT (CIFAR-10) & 69.2 &  0.63  & 0.22 \\

 QViT (MNIST) & 99.5 &  0.02  & 0.02 \\
 \hline
\end{tabular}
\label{tab:Perf}
\end{table*}

\subsubsection{Comparative Analysis of Accuracy and Generalization Error}

\begin{figure}[htbp]
\centerline{\includegraphics[scale=0.36]{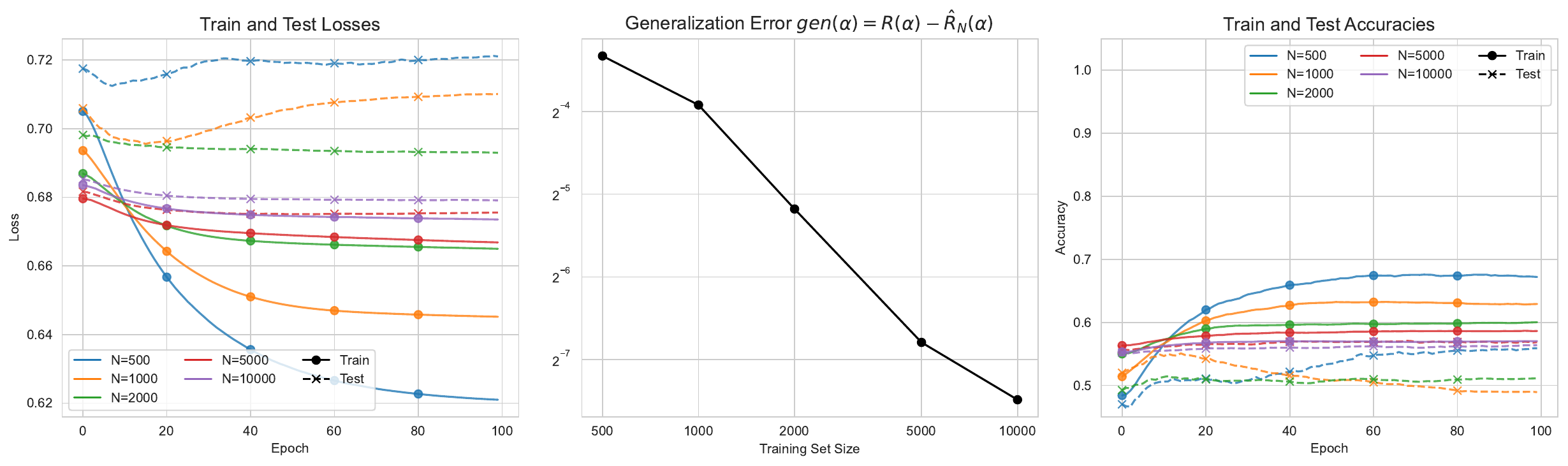}}
\Description{}
\caption{The generalization bound was measured on the Cat vs. Dog pair of the CIFAR-10 dataset for the QRNN model.}
\label{fig:QRNN-Gen}
\end{figure}

The detailed results of the experiment are presented in Table \ref{tab:Perf}. With the MNIST dataset, a low-feature dataset, all three models achieved very high accuracy levels, namely 99.5\% for QViT, 97.3\% for QCNN, and 96.7\% for QRNN. In contrast, the CIFAR-10 dataset, being a high-feature dataset, resulted in lower accuracy for the models. The QViT model achieved the highest accuracy at 69.2\%, while the QCNN model had the lowest at 55.5\%. However, the value returned from the loss function of QCNN was the lowest at 0.48, followed by QViT and QRNN at 0.63 and 0.67, respectively. Thus, the QCNN model is producing approximate probabilities but has not yet been able to translate them into accurate predictions. All three models demonstrated good learning generalization regarding the generalization bound metric, with results of 0.001, 0.013, and 0.22, respectively. We conducted an inspection of the optimization process for all three models. We found no indications of barren plateaus, as shown clearly in the 2D and 3D loss landscape plots in Fig. \ref{fig:landscape} of the QViT model. An interesting observation that can be easily seen from the experimental results with QRNN, as described in Fig. \ref{fig:QRNN-Gen}, is that QNN models can achieve high-fidelity predictions using only a few training data points. The model can already reach its maximum accuracy with just 500 data samples and 100 training epochs. Even when training continues with larger datasets (e.g., 1,000, 2,000, 5,000, or 10,000 samples), the model does not learn further; its learning ability decreases. This result also holds for QCNN and QViT, aligning with previous studies \cite{b15}. All three models replicated in our study are hybrid classical-quantum architectures and utilize the Adam classical optimizer.
\begin{figure}[htbp]
\centerline{\includegraphics[scale=0.4]{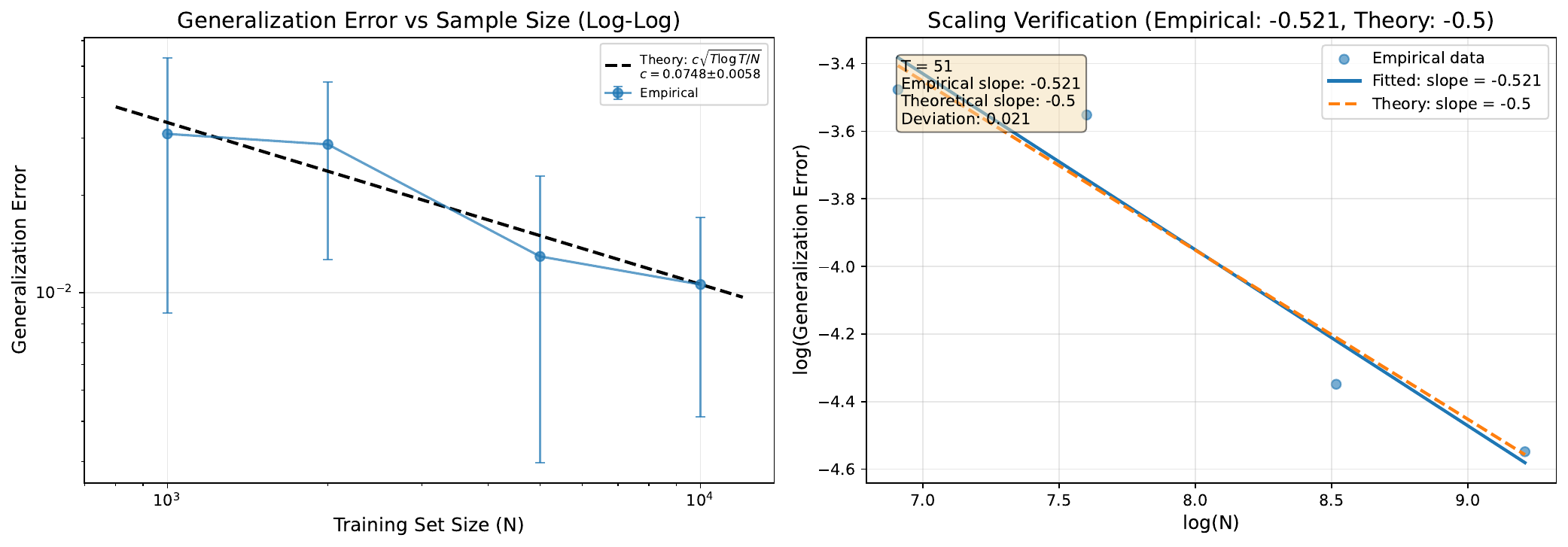}}
\Description{}
\caption{Experimental verification of the claims regarding the alignment between theoretical generalization error and the practical performance of the QCNN model on the CIFAR-10 dataset.}
\label{fig:QCNN-Theo_Prac_Gen}
\end{figure}

To investigate whether the theoretical generalization bound holds in practice, we conducted experiments to examine the relationship between generalization error and training set size. The results show that the test errors for QCNN and QRNN increase in accordance with the prediction $Q\sqrt{T \log T / N}$. In contrast, QViT exhibits an anomaly where the empirical and theoretical curves diverge. Fig. \ref{fig:QCNN-Theo_Prac_Gen} demonstrates the scaling behavior of generalization errors relative to the training set size, while illustrating the interplay between the theoretical upper bounds and the empirical performance of the QCNN model.

\subsubsection{Comparative Analysis of Robustness with Adversarial Noise}

In this section, we provide an in-depth analysis of the robustness characteristics of the evaluated models, leveraging a comprehensive suite of both classical and quantum-centric metrics. Regarding the variability of the Lipschitz constant subjected to adversarial perturbations, based on the results in Table \ref{tab:Lipschitz}, the QRNN model exhibited the highest stability during learning, with an extremely low Lipschitz bound of 0.033, followed by the QCNN model with a value of 0.67. In contrast to the QRNN and QCNN models, the QViT model showed significant instability, with a high Lipschitz bound of 61.38. 

\begin{table*}[htbp]
\centering
\caption{Robustness evaluation of the three models based on the Lipschitz bound across the MNIST and CIFAR-10 datasets.}
\setlength{\tabcolsep}{3pt}
\begin{tabular}{p{50pt}p{120pt}p{100pt}}
    \toprule
    \textbf{Model} & \textbf{Lipschitz Bound (CIFAR-10)} & \textbf{Lipschitz Bound (MNIST)}\\
    \midrule
    QCNN  & 0.67    & 1.09 \\

    QRNN  & 0.033   & 0.12  \\

    QViT  & 61.38   & 68.1 \\
    \hline
\end{tabular}
\label{tab:Lipschitz}
\end{table*}

Table \ref{tab:Adv} presents the prediction results of the models on the CIFAR-10 dataset subjected to adversarial attacks using APGD. The $\epsilon$ parameter describes the intensity of the adversarial matrix attack on the data. For the QCNN model, the accuracy dropped significantly with $\epsilon = 0.1$, from 55.5\% to 31.6\%. However, with higher $\epsilon$ values, the accuracy did not decline sharply, only decreasing by 2–3\%. For the QViT model, the accuracy dropped to 0, indicating that the model has no resistance to the noise generated by the adversarial attack matrix. In contrast, the QRNN model exhibited the best resistance among the three models, with accuracy decreasing from 57.1\% to 47.6\% at $\epsilon = 0.1$, and at $\epsilon = 0.5$, the highest attack intensity, the accuracy only dropped to 45.5\%.

\begin{table*}[htbp]
\caption{Evaluation of the robustness of the QNN models on noisy CIFAR-10 datasets using the APGD attack method at the first epoch.}
\label{tab:Adv}
\setlength{\tabcolsep}{3pt}
\begin{tabular}{p{30pt}p{30pt}p{110pt}p{85pt}}
    \toprule
    \textbf{Model} & \textbf{Epsilon} & \textbf{Adversarial Success Rate (\%)} & \textbf{Average Fidelity} \\
    \midrule
      & 0.1  & 68.4  & 0.73 \\
QCNN  & 0.2  & 70    & 0.71 \\
      & 0.3  & 71.4  & 0.70 \\
      & 0.5  & 73.4  & 0.67 \\
 \hline
      & 0.1  & 52.4 & 0.671\\
QRNN  & 0.2  & 52.9 & 0.670\\ 
      & 0.3  & 53.5 & 0.669\\ 
      & 0.5  & 54.5 & 0.668\\
 \hline
      & 0.1  & 100 & 0.86 \\
QViT  & 0.2  & 100 & 0.72 \\
      & 0.3  & 100 & 0.61 \\
      & 0.5  & 100 & 0.45 \\
 \hline
\end{tabular}
\end{table*}

Experimental research on QNN models reveals an adversarial relationship between accuracy and resilience to adversarial attacks. The QViT model achieved the highest accuracy at 68.3\%; however, its adversarial accuracy dropped to 0\% even at the smallest perturbation level ($\epsilon=0.1$). In contrast, while the QRNN and QCNN models did not exhibit learning capabilities as strong as QViT, they demonstrated superior resistance to adversarial attacks, with a robustness gap of approximately 25\% for QCNN and 3\% for QRNN. These findings indicate that the superior learning performance of QViT comes at the cost of heightened vulnerability to attacks. Conversely, the modest accuracy of QRNN and QCNN enables them to remain less affected by changes in input images. This adversarial relationship can be attributed to the architectural characteristics of QNN models.

The use of many parameters and the global quantum self-attention mechanism may contribute to the superior learning capability, but poor resilience to adversarial attacks of the QViT model. Additionally, the global self-attention mechanism causes each patch to attend to all patches in the image, resulting in a slight change that potentially affects the output of all patches. These characteristics of QViT are evidenced by its considerable Lipschitz bound value (118). The QViT model tends to become more sensitive to changes in input images as it is trained for more epochs. For both CIFAR and MNIST datasets, and across all class pairs, the model's empirical Lipschitz constant (LC) tends to increase with the number of training epochs. Training a model is essentially a process of minimizing the loss function; in other words, it pushes the predicted probabilities closer to 0 for negative samples and closer to 1 for positive samples. Since QViT outputs pre-softmax logits, the model drives these logits toward extreme values, increasing the magnitude of the input-output Jacobian. This increasing magnitude could explain why the Lipschitz bound, computed as the maximum norm of the gradient of the model's output concerning the input across training samples,positively correlates with the number of epochs. The LC will likely keep increasing as long as the model has sufficient capacity to fit the training set.

\begin{figure}[htbp]
\centerline{\includegraphics[scale=0.35]{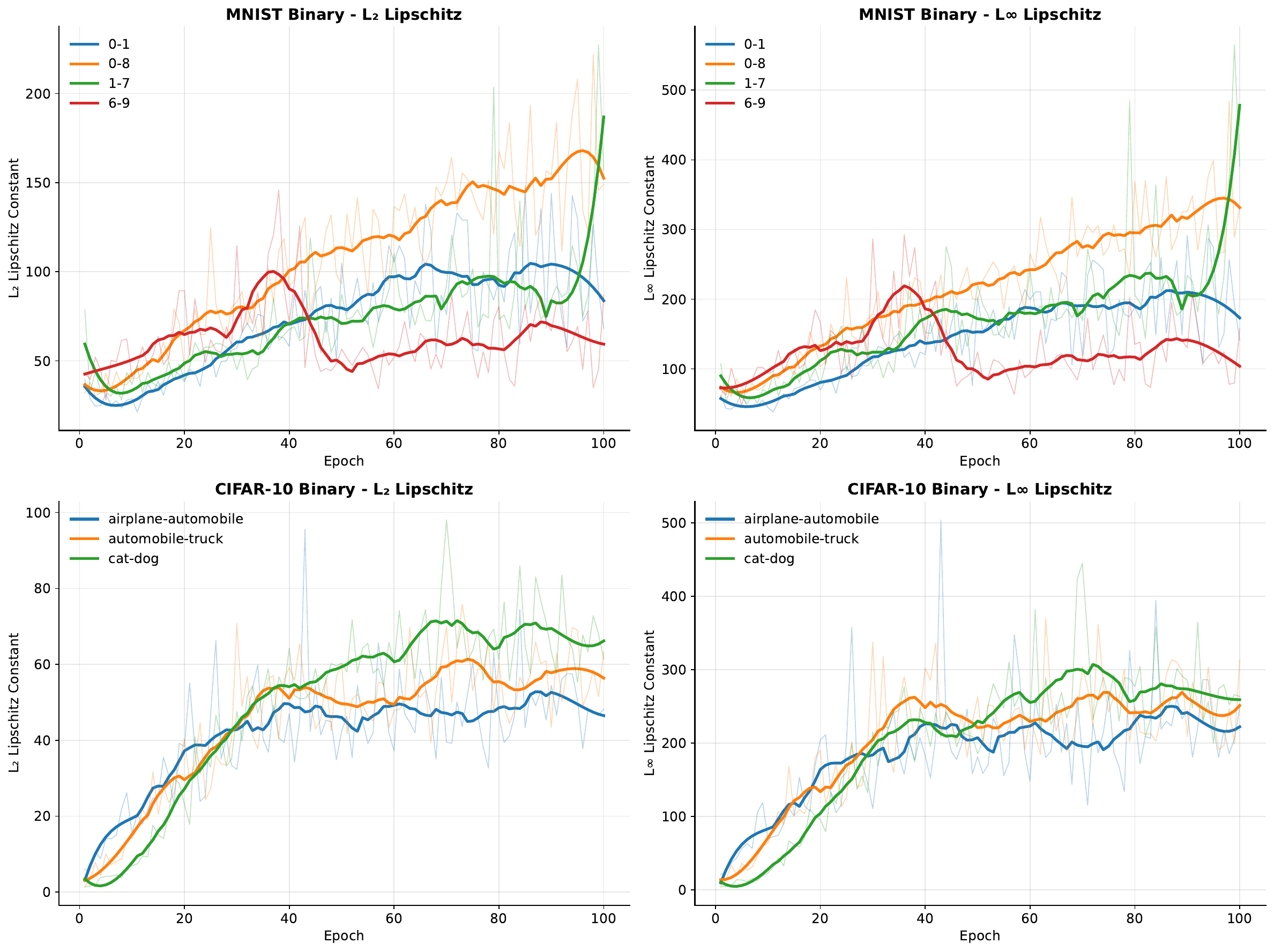}}
\Description{}
\caption{The measurement results of the variation in Lipschitz bound across training epochs on the MNIST and CIFAR-10 datasets for the binary classification task, with two noise bounds, $L_{2}$ and $L_{\infty}$, for the QViT model.  a) and b) The variation of the Lipschitz bound across epochs in the binary classification task, where 0-1, 0-8, 1-7, 6-9 are the handwritten digit pairs from the MNIST dataset. c) and d) Graphs depicting the variation of the Lipschitz bound function across epochs, where 0-1, 1-9, 3-5 are the airplane-automobile, automobile-truck, cat-dog pairs from the CIFAR-10 dataset.}
\label{fig:Lips}
\end{figure}

Moreover, Fig. \ref{fig:Lips} also shows that the LC is dataset-dependent. The MNIST dataset generally exhibits a higher Lipschitz bound than CIFAR-10 for $L_{2}$ and $L_{\infty}$. This result may be because class separation in MNIST primarily relies on low-level features (such as edges and curves), making the model more sensitive to small perturbations in the input images. Many parameters enable the model to fit nicely on the training set, achieving high performance on the test set. However, overparameterization and the global self-attention mechanism create a highly complex decision boundary, making the model highly sensitive to small input changes. In contrast to the complexity and global mechanism of QViT, the simplicity of parameters and the local-sequential mechanisms of QCNN and QRNN enable them to resist adversarial attacks effectively.
The robustness against adversarial attacks of QRNN and QCNN can be explained by their architectures' local and sequential nature. The QCNN model, similar to traditional CNNs, employs convolutional and pooling layers to process inputs locally. Meanwhile, QRNN, with the lowest Lipschitz bound, extracts features sequentially. Concurrently, QRNN achieves the best adversarial success rate (ASR) of 52.4\% and a robustness gap of 3\%. The local nature of QCNN reduces the impact of local changes on the information of the entire image. The QRNN model generates representations sequentially, resulting in a smoother decision boundary and enhancing resilience to small input changes. 


\begin{figure}[htbp]
    \centerline{\includegraphics[scale=0.22]{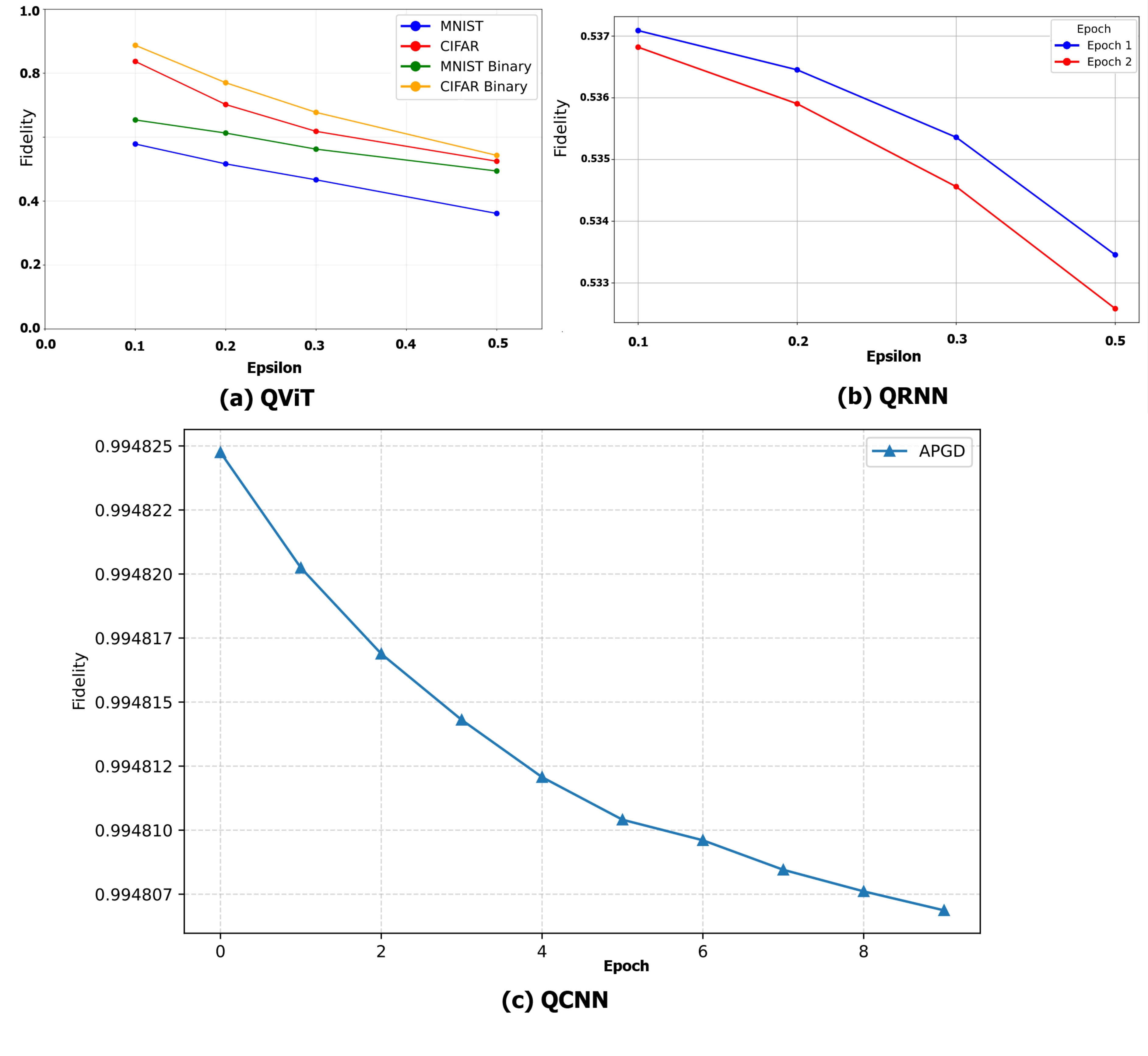}}
    \Description{}
    \caption{\large Comparison of Average Fidelity across QNN models.}
    \Description{}
    \label{fig:comparison}
\end{figure}


For quantum metrics, the average fidelity is used to measure the similarity between the quantum state of the image before and after the attack. It can be observed that after the attack, the quantum state changes differently across the models. In image a), QCNN demonstrates relatively good resilience against adversarial attacks, as the average fidelity value decreases only slightly with the increase in epochs under the strongest attack method, APGD. For the QRNN model, which is the most resistant to attacks, the image state shows minimal changes across different attack intensities, with a slow decrease in fidelity from 0.671 to 0.668. Conversely, for the QViT model, which is the least resistant to attacks, fidelity drops sharply from 0.86 at $\epsilon = 0.1$ to 0.45 at $\epsilon = 0.5$, indicating that the data is heavily attacked and is a significant factor in the model's inability to predict the correct class and its weak resistance to attacks. The trends in average fidelity variation across the three models are clearly visible in Fig. \ref{fig:comparison}. In summary, we have the following observations from experiments on QNN models: QViT exhibits the lowest degradation, possibly due to its quantum attention mechanism and nonlinear encoding. QRNN generalizes better on temporal or sequential datasets. QNN models learn effectively and show natural potential in resisting adversarial attacks.

\subsubsection{Comparative Analysis of Robustness with Quantum Noise}

\begin{figure}[htbp]
\centerline{\includegraphics[scale=0.4]{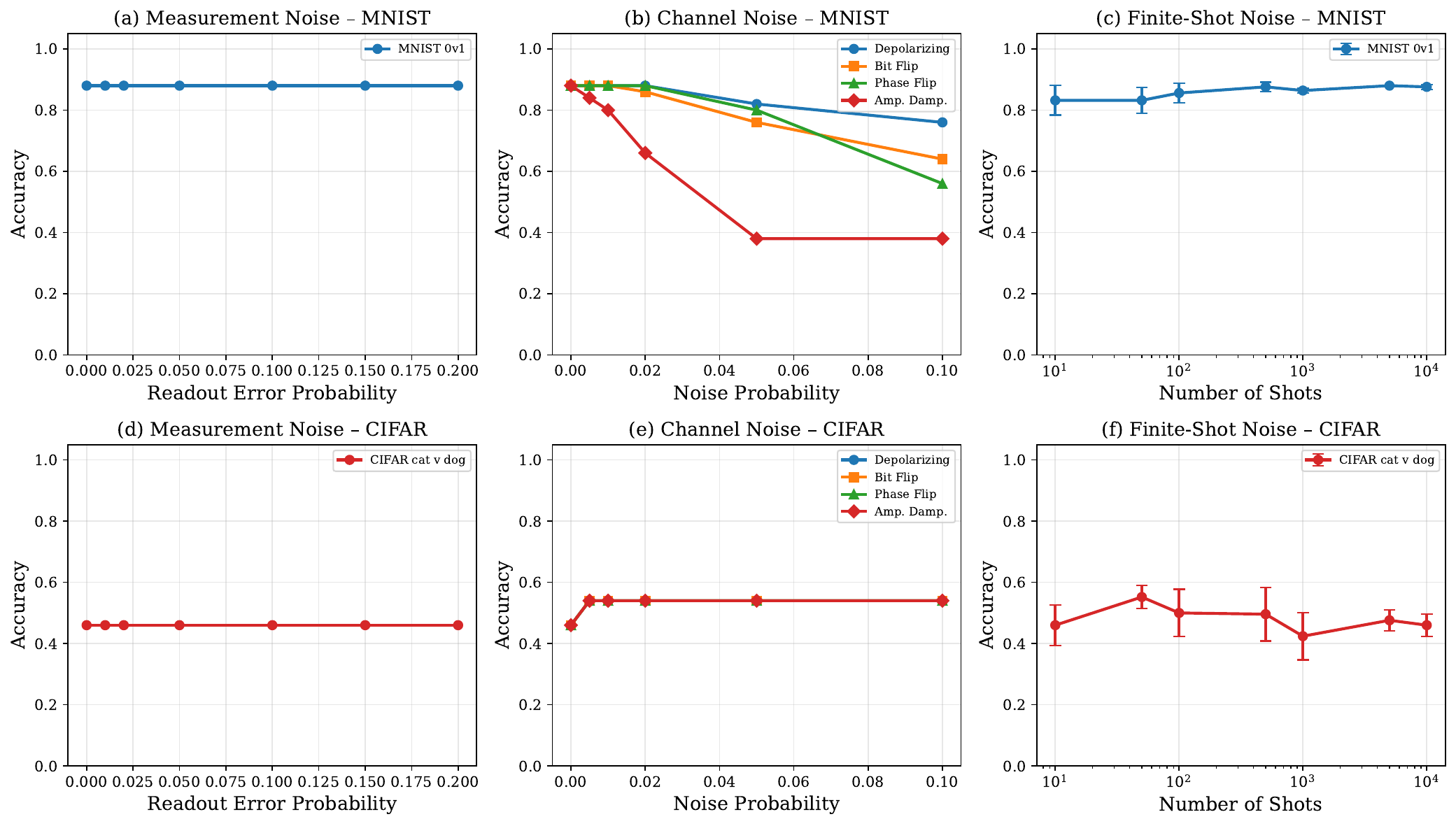}}
\Description{}
\caption{Impact of quantum noise on the performance of the QRNN model.}
\label{fig:quantum_noise}
\end{figure}

To provide an objective assessment of the robustness of the QNN models, we selected quantum noise as the evaluation metric. We utilized measurement and channel noise types and examined the impact of finite-shot effects on the accuracy of these models. The results indicate that the robustness of the QCNN architecture was rigorously evaluated across five quantum noise channels, revealing a common threshold at which performance typically collapses at a noise probability of 0.5. For Amplitude-damping and Bit-flip noise, QCNNs demonstrated significant resilience at lower intensities, occasionally exhibiting 'noise-enhanced learning' in which training accuracy exceeded the ideal, though without improving generalization. In contrast, the model proved highly sensitive to Depolarization channel noise, with a sharp decline in learning ability as the probability exceeded 0.2, due to rapid loss of quantum coherence. Interestingly, QCNNs displayed exceptional robustness against Phase-damping and Phase-flip noise, maintaining or even exceeding baseline performance across a broader range of noise levels. While higher noise intensities generally led to a failure to learn, a notable anomaly occurred at a Phase-flip probability of 1.0, where the model regained its ability to adapt. These findings suggest that while QCNNs can effectively internalize certain noise patterns as features, their operational viability remains strictly dependent on the specific noise type and intensity. Ultimately, these results underscore the necessity for tailored noise management strategies when deploying QCNNs on NISQ-era hardware. Experiments conducted on QViT did not reveal a clear or significant impact of quantum noise on the model's performance; consequently, further in-depth empirical evaluations are required to reach definitive conclusions. Regarding the QRNN model, our results indicate that accuracy remains largely unaffected by measurement noise across both MNIST and CIFAR-10 datasets. Conversely, various channel noise types lead to a noticeable degradation in model accuracy which is particularly evident with Amplitude-damping noise on the MNIST dataset, where accuracy drops to approximately 0.4 and subsequently plateaus. Similar to the observations for measurement noise, the finite-shot effect does not significantly affect accuracy regardless of the number of shots used. These results suggest that while the QRNN architecture is resilient to certain statistical fluctuations, it remains highly vulnerable to specific physical decoherence channels. The detailed results regarding the impact of various quantum noise channels on the QRNN model are extensively illustrated in Fig. \ref{fig:quantum_noise}.

\subsubsection{Discussion}

The research results in this paper can be applied to various other fields. In addition to traditional applications, such as natural language processing, image classification, and object detection, QNN models can be modified for healthcare applications, such as using QNNs to predict patient coma levels from EEG indices. Furthermore, in the field of physical simulation, QNNs can improve the ability to solve stiff, highly nonlinear partial differential equations. Some of our current research is focusing on these potential directions. The minor variants of the VQC algorithm used in these QNN models include Quantum Approximate Optimization Algorithms (QAOA) for general optimization problems, Quantum Support Vector Machine (QSVM) for discrete data prediction problems, Quantum Physics-Informed Neural Networks (QPINN) for solving differential equations in physics, Quantum Deep Reinforcement Learning (QRL) for robotics, and Quantum Federated Learning (QFL) for distributed machine learning problems.

Based on the experimental results, we propose several solutions to improve the performance of hybrid quantum-classical QNN models. First, limit the depth of the main VQC quantum circuits to reduce the impact of noise on the model's prediction quality and to accommodate the simulation capabilities of current classical computers while quantum computers are still in the NISQ era. Second, focus on exploiting QNN's strengths for low-feature datasets \cite{b10}, or on transforming local data with a small number of qubits to fit the Hilbert space of NISQ machines \cite{tran2025quantum}. Third, use trainable embedding methods; i.e., machine learning models can be employed to improve embedding quality which has been proven to enhance model performance significantly. Additionally, regularization techniques can help improve QNN models, as they share properties with classical DL models.


\section{Conclusions and Future Research}

In this study, we conducted a series of detailed experiments to evaluate comparatively three fundamental QNN models,QCNN, QRNN, and QViT, across two dimensions: generalization and robustness. We employed key classical machine learning metrics, including generalization error, accuracy, loss function, and Lipschitz bound, as well as a quantum evaluation metric, average fidelity, to benchmark these three models. The experimental results indicate that QViT achieves the best performance in terms of generalization but is prone to overfitting as the number of training epochs increases. Using an excessive number of parameters to achieve higher accuracy compared to other models also leads to significant hardware resource consumption for QViT. Although QRNN is not highly regarded in theory for its image processing capabilities, it demonstrates higher accuracy than QCNN, a model specifically designed for this domain. Regarding robustness, QRNN demonstrated the highest resilience among the three models. In contrast, QViT is highly sensitive to adversarial examples and the most vulnerable to attacks compared to the other two models. Additionally, APGD proved to be the most effective attack method against QNN models, consistently achieving high success rates across all noise levels. Among the three quantum deep learning models we researched, two models, QRNN and QViT, have a sequence-to-sequence architecture. The other model, QCNN, is a convolutional model with a filter window scanning the entire lattice space. While QRNN and QViT are inherently sequence-based models, QViT employs an attention mechanism, resulting in significantly higher parameter requirements than QRNN. These factors contribute to the performance characteristics of the models, as detailed in the experimental results section. 

This study has provided insights into the learning capabilities and resilience against adversarial attacks of QNN models; however, there are still limitations for future research to explore. The study is limited to binary classification for the QRNN and QCNN models, and the hyperparameter space has not been thoroughly investigated. The trade-offs and behaviors of the models discussed above may vary for more complex tasks, such as multi-class classification, or with different sets of hyperparameters. The benchmarking has been limited to image classification tasks and has not yet addressed other deep learning problems, such as image detection, real-time video processing, or natural language processing. Therefore, future research should focus on exploring the characteristics of QNN models in more complex tasks while identifying configurations that enable these models to achieve optimal performance. Our study provides researchers with a deeper insight into the performance of current QNN models. It supports further research on selecting and applying these models to image classification tasks in specific domains.

In the next step of our research, we will conduct further experiments to explore the impact of quantum noise on model accuracy on real quantum computers. We will also investigate pure quantum optimizers to understand the relationship between parameter-shift optimization and the barren plateau problem. Furthermore, the influence of embedding methods will be more broadly evaluated with other types of embeddings, rather than being limited to just the two main methods: angle and amplitude.

\section{Data Sources}
The datasets utilized in this study are available for access here: \href{https://huggingface.co/datasets/ylecun/mnist}{MNIST} - \href{https://huggingface.co/datasets/uoft-cs/cifar10}{CIFAR-10}.

\bibliographystyle{ACM-Reference-Format}
\bibliography{QCAM}

@String{Computing = "Computing" }

@String{Computer = "{IEEE} Computer" }

@String{Springer = "Springer-Verlag" }

@article{b1,
  title={Quantum machine learning},
  author={Biamonte, Jacob and Wittek, Peter and Pancotti, Nicola and Rebentrost, Patrick and Wiebe, Nathan and Lloyd, Seth},
  journal={Nature},
  volume={549},
  number={7671},
  pages={195--202},
  year={2017},
  doi={10.1038/nature23474}
}

@article{b2,
  title={Deep learning},
  author={LeCun, Yann and Bengio, Yoshua and Hinton, Geoffrey},
  journal={nature},
  volume={521},
  number={7553},
  pages={436--444},
  year={2015},
  publisher={Nature Publishing Group UK London}
}

@article{b4,
  title={Quantum deep learning},
  author={Wiebe, Nathan and Kapoor, Alireza and Svore, Krysta M},
  journal={Quantum Information and Computation},
  volume={16},
  number={7-8},
  pages={541--587},
  year={2016},
  doi={10.26421/qic16.7-8-1}
}

@article{b5,
  title={Variational quantum algorithms},
  author={Cerezo, Marco and Arrasmith, Andrew and Babbush, Ryan and Benjamin, Simon C and Endo, Suguru and Fujii, Keisuke and McClean, Jarrod R and Mitarai, Kosuke and Yuan, Xiao and Cincio, Lukasz and others},
  journal={Nature Reviews Physics},
  volume={3},
  number={9},
  pages={625--644},
  year={2021},
  publisher={Nature Publishing Group UK London}
}

@article{b10,
  title={Quanvolutional neural networks: powering image recognition with quantum circuits},
  author={Henderson, Maxwell and Shakya, Samriddhi and Pradhan, Shashindra and Cook, Tristan},
  journal={Quantum Machine Intelligence},
  volume={2},
  number={1},
  pages={2},
  year={2020},
  doi={10.1007/s42484-020-00012-y}
}

@article{b11,
  title={Quantum convolutional neural networks},
  author={Cong, Iris and Choi, Soonwon and Lukin, Mikhail D},
  journal={Nature Physics},
  volume={15},
  number={12},
  pages={1273--1278},
  year={2019},
  doi={10.1038/s41567-019-0648-8}
}

@article{b12,
  title={Quantum recurrent neural networks for sequential learning},
  author={Li, Yanan and others},
  journal={Neural Networks},
  volume={166},
  pages={148--161},
  year={2023},
  doi={10.1016/j.neunet.2023.07.003}
}

@article{b13,
  title={Quantum vision transformers},
  author={Cherrat, El Amine and Kerenidis, Iordanis and Mathur, Natansh and Landman, Jonas and Strahm, Martin and Li, Yun Yvonna},
  journal={Quantum},
  volume={8},
  pages={1265},
  year={2024},
  doi={10.22331/q-2024-02-22-1265}
}

@article{b15,
  title={Generalization in quantum machine learning from few training data},
  author={Caro, Matthias C and others},
  journal={Nature Communications},
  volume={13},
  number={1},
  pages={4919},
  year={2022},
  doi={10.1038/s41467-022-32550-3}
}

@article{b16,
  title={Quantum adversarial machine learning},
  author={Lu, Sirui and Duan, Lu-Ming and Deng, Dong-Ling},
  journal={Physical Review Research},
  volume={2},
  number={3},
  pages={033212},
  year={2020},
  doi={10.1103/physrevresearch.2.033212}
}

@article{b18,
  title={Class of quantum many-body states that can be efficiently simulated},
  author={Vidal, Guifre},
  journal={Physical Review Letters},
  volume={101},
  number={11},
  pages={110501},
  year={2008},
  doi={10.1103/physrevlett.101.110501}
}

@article{b19,
  title   = {PennyLane: Automatic differentiation of hybrid quantum-classical computations},
  author  = {Bergholm, Ville and Izaac, Josh and Schuld, Maria and Gogolin, Christian and Ahmed, Shahnawaz and Ajith, Vishnu and Alam, M. Sohaib and Alonso-Linaje, Guillermo and AkashNarayanan, B. and Asadi, Ali and others},
  journal = {arXiv preprint},
  volume  = {abs/1811.04968},
  number  = {1811.04968},
  year    = {2018},
  pages   = {1--19},
  note    = {arXiv:1811.04968}
}

@article{b20,
  title   = {Recurrent Neural Networks ({RNNs}): A Gentle Introduction and Overview},
  author  = {Schmidt, Richard M.},
  journal = {arXiv preprint},
  volume  = {abs/1912.05911},
  number  = {1912.05911},
  year    = {2019},
  pages   = {1--40},
  note    = {arXiv:1912.05911}
}

@inproceedings{b21,
  title     = {Recurrent Quantum Neural Networks},
  author    = {Bausch, Johannes},
  booktitle = {Advances in Neural Information Processing Systems (NeurIPS)},
  volume    = {33},
  pages     = {1368--1379},
  year      = {2020},
  publisher = {Curran Associates, Inc.},
  address   = {Red Hook, NY, USA}
}

@article{b22,
  title   = {An Image is Worth 16x16 Words: Transformers for Image Recognition at Scale},
  author  = {Dosovitskiy, Alexey and Beyer, Lucas and Kolesnikov, Alexander and Weissenborn, Dirk and Zhai, Xiaohua and Unterthiner, Thomas and Dehghani, Mostafa and Minderer, Matthias and Heigold, Georg and Gelly, Sylvain and others},
  journal = {arXiv preprint},
  volume  = {abs/2010.11929},
  number  = {2010.11929},
  year    = {2021},
  pages   = {1--22},
  note    = {arXiv:2010.11929}
}

@article{b23,
  title={Learning temporal data with a variational quantum recurrent neural network},
  author={Takaki, Yoshiki and Mitarai, Kosuke and Negoro, Makoto and Fujii, Keisuke and Kitagawa, Masahiro},
  journal={Physical Review A},
  volume={103},
  number={5},
  pages={052414},
  year={2021},
  doi={10.1103/physreva.103.052414}
}

@article{b24,
  title   = {Attention Is All You Need},
  author  = {Vaswani, Ashish and Shazeer, Noam and Parmar, Niki and Uszkoreit, Jakob and Jones, Llion and Gomez, Aidan N. and Kaiser, {\L}ukasz and Polosukhin, Illia},
  journal = {arXiv preprint},
  volume  = {abs/1706.03762},
  number  = {1706.03762},
  year    = {2017},
  pages   = {1--15},
  note    = {arXiv:1706.03762}
}

@inproceedings{b25,
  title     = {An Image is Worth 16x16 Words: Transformers for Image Recognition at Scale},
  author    = {Dosovitskiy, Alexey and Beyer, Lucas and Kolesnikov, Alexander and Weissenborn, Dirk and Zhai, Xiaohua and Unterthiner, Thomas and Dehghani, Mostafa and Minderer, Matthias and Heigold, Georg and Gelly, Sylvain and others},
  booktitle = {International Conference on Learning Representations (ICLR)},
  year      = {2021},
  pages     = {1--21},
  publisher = {OpenReview.net},
  address   = {Vienna, Austria}
}

@inproceedings{b26,
  title     = {The Dawn of Quantum Natural Language Processing},
  author    = {Di Sipio, Riccardo and Huang, Jiun-Hung and Chen, Shih-Yuan C. and Mangini, Stefano and Worring, Marcel},
  booktitle = {2022 IEEE International Conference on Acoustics, Speech and Signal Processing (ICASSP)},
  pages     = {8612--8616},
  year      = {2022},
  doi       = {10.1109/icassp43922.2022.9747675},
  publisher = {IEEE},
  address   = {Singapore}
}

@article{b27,
  title={Quantum self-attention neural networks for text classification},
  author={Li, Gang and Zhao, Xiaoliang and Wang, Xiugang},
  journal={Science China Information Sciences},
  volume={67},
  number={4},
  pages={142501},
  year={2024},
  doi={10.1007/s11432-023-3879-7}
}

@article{b28,
  title={The nature of statistical learning theory},
  author={Sain, Susan R and Vapnik, Vladimir N},
  journal={Technometrics},
  volume={38},
  number={4},
  pages={409},
  year={1996},
  doi={10.2307/1271324}
}

@article{b29,
  title={Challenges and opportunities in quantum machine learning},
  author={Cerezo, Marco and Verdon, Guillaume and Huang, Hsin-Yuan and Cincio, Lukasz and Coles, Patrick J},
  journal={Nature computational science},
  volume={2},
  number={9},
  pages={567--576},
  year={2022},
  publisher={Nature Publishing Group US New York}
}

@book{b30,
  title     = {An Introduction to Error Analysis: The Study of Uncertainties in Physical Measurements},
  author    = {Taylor, John R.},
  edition   = {2},
  publisher = {University Science Books},
  address   = {Sausalito, CA, USA},
  year      = {1996}
}

@inproceedings{b31,
  title     = {Objective functions for neural network classifier design},
  author    = {Goodman, R. M. and Miller, J. W. and Smyth, P.},
  booktitle = {Proceedings of 1991 IEEE International Symposium on Information Theory},
  pages     = {87--87},
  year      = {1991},
  publisher = {IEEE},
  address   = {Budapest, Hungary},
  doi       = {10.1109/ISIT.1991.695123}
}

@article{b32,
  title   = {Loss Functions and Metrics in Deep Learning},
  author  = {Terven, Juan and Cordova-Esparza, Daniel M. and Ramirez-Pedraza, Alan and Chavez-Urbiola, Esthela A. and Romero-Gonzalez, Jose A.},
  journal = {arXiv preprint},
  volume  = {abs/2307.02694},
  number  = {2307.02694},
  year    = {2023},
  pages   = {1--35},
  note    = {arXiv:2307.02694}
}

@article{b34,
  title={Training robust and generalizable quantum models},
  author={Berberich, Julian and Fink, Daniel and Pranji{\'c}, Daniel and Tutschku, Christian and Holm, Christian},
  journal={Physical Review Research},
  volume={6},
  number={4},
  pages={043326},
  year={2024},
  doi={10.1103/physrevresearch.6.043326}
}

@article{b35,
  title={A simple formula for the average gate fidelity of a quantum dynamical operation},
  author={Nielsen, Michael A},
  journal={Physics Letters A},
  volume={303},
  number={4},
  pages={249--252},
  year={2002},
  doi={10.1016/s0375-9601(02)01272-0}
}

@article{b42,
  title   = {The {MNIST} handwritten digit database},
  author  = {LeCun, Yann and Cortes, Corinna and Burges, Christopher J. C.},
  journal = {AT\&T Labs [Online]},
  volume  = {2},
  number  = {5},
  year    = {2010},
  pages   = {1--2},
  url     = {http://yann.lecun.com/exdb/mnist/},
  note    = {Available at http://yann.lecun.com/exdb/mnist/}
}

@misc{b43,
  title={Learning multiple layers of features from tiny images},
  author={Krizhevsky, Alex},
  howpublished={Tech. Rep., Univ. Toronto},
  year={2009},
  url={https://www.cs.toronto.edu/~kriz/cifar.html}
}

@inproceedings{b44,
  title     = {Explaining and Harnessing Adversarial Examples},
  author    = {Goodfellow, Ian J. and Shlens, Jonathon and Szegedy, Christian},
  booktitle = {3rd International Conference on Learning Representations (ICLR)},
  year      = {2015},
  pages     = {1--11},
  publisher = {OpenReview.net},
  address   = {San Diego, CA, USA}
}

@inproceedings{b45,
  title     = {Towards Deep Learning Models Resistant to Adversarial Attacks},
  author    = {Madry, Aleksander and Makelov, Aleksandar and Schmidt, Ludwig and Tsipras, Dimitris and Vladu, Adrian},
  booktitle = {6th International Conference on Learning Representations (ICLR)},
  year      = {2018},
  pages     = {1--28},
  publisher = {OpenReview.net},
  address   = {Vancouver, BC, Canada}
}

@inproceedings{b46,
  title     = {Reliable evaluation of adversarial robustness with an ensemble of diverse parameter-free attacks},
  author    = {Croce, Francesco and Hein, Matthias},
  booktitle = {Proceedings of the 37th International Conference on Machine Learning (ICML)},
  series    = {Proceedings of Machine Learning Research},
  volume    = {119},
  pages     = {2206--2216},
  year      = {2020},
  publisher = {PMLR},
  address   = {Vienna, Austria}
}

@inproceedings{b47,
  title     = {Boosting Adversarial Attacks with Momentum},
  author    = {Dong, Yinpeng and Liao, Fangzhou and Pang, Tianyu and Su, Hang and Zhu, Jun and Hu, Xiaolin and Li, Jianguo},
  booktitle = {Proceedings of the IEEE/CVF Conference on Computer Vision and Pattern Recognition (CVPR)},
  pages     = {9185--9193},
  year      = {2018},
  publisher = {IEEE},
  address   = {Salt Lake City, UT, USA}
}

@inproceedings{b49,
  title     = {A survey of classical and quantum sequence models},
  author    = {Chen, I-Chung and Singh, Harmeet and Anukruti, V. L. and Quanz, Beate and Yogaraj, Kavitha},
  booktitle = {2024 16th International Conference on Communication Systems and Networks (COMSNETS)},
  pages     = {1006--1011},
  year      = {2024},
  doi       = {10.1109/comsnets59351.2024.10456721},
  publisher = {IEEE},
  address   = {Bengaluru, India}
}

@article{b55,
  title   = {Classification with Quantum Neural Networks on Near Term Processors},
  author  = {Farhi, Edward and Neven, Hartmut},
  journal = {arXiv preprint},
  volume  = {abs/1802.06002},
  number  = {1802.06002},
  year    = {2018},
  pages   = {1--21},
  note    = {arXiv:1802.06002}
}

@article{ahmed2025comparative,
  title={A comparative analysis and noise robustness evaluation in quantum neural networks},
  author={Ahmed, Tasnim and Kashif, Muhammad and Marchisio, Alberto and Shafique, Muhammad},
  journal={Scientific Reports},
  volume={15},
  number={1},
  pages={33654},
  year={2025},
  publisher={Nature Publishing Group UK London}
}

@inproceedings{zaman2024comparative,
  title={A comparative analysis of hybrid-quantum classical neural networks},
  author={Zaman, Kamila and Ahmed, Tasnim and Hanif, Muhammad Abdullah and Marchisio, Alberto and Shafique, Muhammad},
  booktitle={World Congress in Computer Science, Computer Engineering \& Applied Computing},
  pages={102--115},
  year={2024},
  organization={Springer}
}

@inproceedings{tran2025quantum,
  title        = {Quantum Patches for Efficient Learning},
  author       = {Tran, Ban Q. and Luong, Chuong K. and Mengel, Susan},
  booktitle    = {International Conference on Multi-disciplinary Trends in Artificial Intelligence (MIWAI)},
  pages        = {87--100},
  year         = {2025},
  publisher    = {Springer Nature},
  organization = {Springer},
  address      = {Cham, Switzerland}
}

@article{bowles2024better,
  title={Better than classical? the subtle art of benchmarking quantum machine learning models},
  author={Bowles, Joseph and Ahmed, Shahnawaz and Schuld, Maria},
  journal={arXiv preprint arXiv:2403.07059},
  year={2024}
}

@article{abbas2021power,
  title={The power of quantum neural networks},
  author={Abbas, Amira and Sutter, David and Zoufal, Christa and Lucchi, Aur{\'e}lien and Figalli, Alessio and Woerner, Stefan},
  journal={Nature computational science},
  volume={1},
  number={6},
  pages={403--409},
  year={2021},
  publisher={Nature Publishing Group US New York}
}

@article{schuld2015introduction,
  title={An introduction to quantum machine learning},
  author={Schuld, Maria and Sinayskiy, Ilya and Petruccione, Francesco},
  journal={Contemporary Physics},
  volume={56},
  number={2},
  pages={172--185},
  year={2015},
  publisher={Taylor \& Francis}
}

@article{huang2021power,
  title={Power of data in quantum machine learning},
  author={Huang, Hsin-Yuan and Broughton, Michael and Mohseni, Masoud and Babbush, Ryan and Boixo, Sergio and Neven, Hartmut and McClean, Jarrod R},
  journal={Nature communications},
  volume={12},
  number={1},
  pages={2631},
  year={2021},
  publisher={Nature Publishing Group UK London}
}

@article{schuld2019quantum,
  title={Quantum machine learning in feature Hilbert spaces},
  author={Schuld, Maria and Killoran, Nathan},
  journal={Physical review letters},
  volume={122},
  number={4},
  pages={040504},
  year={2019},
  publisher={APS}
}

@article{ciliberto2018quantum,
  title={Quantum machine learning: a classical perspective},
  author={Ciliberto, Carlo and Herbster, Mark and Ialongo, Alessandro Davide and Pontil, Massimiliano and Rocchetto, Andrea and Severini, Simone and Wossnig, Leonard},
  journal={Proceedings of the Royal Society A: Mathematical, Physical and Engineering Sciences},
  volume={474},
  number={2209},
  pages={20170551},
  year={2018},
  publisher={The Royal Society Publishing}
}

@article{dunjko2018machine,
  title={Machine learning and artificial intelligence in the quantum domain: a review of recent progress},
  author={Dunjko, Vedran and Briegel, Hans J},
  journal={Reports on Progress in Physics},
  volume={81},
  number={7},
  pages={074001},
  year={2018},
  publisher={IOP Publishing}
}

@article{basilewitsch2025quantum,
  title={Quantum neural networks in practice: a comparative study with classical models from standard data sets to industrial images},
  author={Basilewitsch, Daniel and Bravo, Jo{\~a}o F and Tutschku, Christian and Struckmeier, Frederick},
  journal={Quantum Machine Intelligence},
  volume={7},
  number={2},
  pages={110},
  year={2025},
  publisher={Springer}
}

\end{document}